\documentclass[reprint,twocolumn,pre,amsmath,amssymb,amsfonts,aps,floatfix]{revtex4-2}

\usepackage{natbib}
\usepackage{graphicx,color,rotating}
\usepackage[latin1]{inputenc}
\usepackage{textcomp}
\usepackage{dcolumn}
\usepackage{bm}     
\usepackage{upgreek}
\usepackage{xfrac}
\usepackage{array} 
\newcolumntype{L}[1]{>{\raggedright\let\newline\\\arraybackslash\hspace{0pt}}m{#1}}
\newcolumntype{C}[1]{>{\centering\let\newline\\\arraybackslash\hspace{0pt}}m{#1}}
\newcolumntype{R}[1]{>{\raggedleft\let\newline\\\arraybackslash\hspace{0pt}}m{#1}}

\usepackage{hyperref}					
\usepackage{xcolor}						
\hypersetup{			     			
    colorlinks,
    allcolors={blue}
}
\newcommand{\qm}[1]{{``#1''}}
\newcommand{\qmtt}[1]{{``\texttt{#1}''}}
\newcommand{\qmttL}[1]{{``\texttt{#1}}}
\newcommand{\qmttC}[1]{{\texttt{#1}}}
\newcommand{\qmttR}[1]{{\texttt{#1}''}}

\newcommand{\upspace}{\rule{0ex}{2.5ex}}
\newcommand{\downspace}{\rule[-1.5ex]{0ex}{1.0ex}}
\newcommand{\negspace}{\hspace*{-2mm}}
\newcommand{\notop}{{{}_{}}}

\newcommand{\edbLM}[2]{\makebox[#1ex]{\edb{\rule[0.2ex]{#1ex}{0.1ex}}}\hspace*{-#1ex}\makebox[#1ex]{$\dpst#2$}}

\newcommand{\timeav}[2]{\big\langle #1, #2 \big\rangle}

\newcommand{\mr}[1]{\ensuremath{\mathrm{#1}}}
\newcommand{\myvec}[1]{\bm{#1}}
\newcommand{\ee}{\mathrm{e}}
\newcommand{\ii}{\mathrm{i}}
\newcommand{\dm}{\mathrm{d}}

\newcommand{\avr}[1]{\big\langle #1 \big\rangle}

\newcommand{\intnab}[3]{I^{(#1)}_{{#2}{#3}}}
\newcommand{\intBCnab}[3]{\hat{I}^{(#1)}_{{#2}{#3}}}

\newcommand{\re}{\mathrm{Re}}

\newcommand{\ve}{\varepsilon}

\newcommand{\pp}{\partial^{{}}}
\newcommand{\ppsqr}{\partial^{\,2_{}}}

\newcommand{\ppt}{\pp_t}

\newcommand{\nablabf}{\boldsymbol{\nabla}}

\newcommand{\Lapl}{\nabla^2}

\newcommand{\grad}{\nablabf}
\newcommand{\curl}{\nablabf\times}
\renewcommand{\div}{\nablabf\cdot}
\newcommand{\divop}{\nablabf\cdot}

\newcommand{\norm}[1]{\left\lVert#1\right\rVert}
\newcommand{\eg}{\textit{e.g.}}

\newcommand{\etal}{\textit{et~al.}}

\newcommand{\scap}{\!\cdot\!}

\newcommand{\dpst}{\displaystyle}

\newcommand{\AAA}{\myvec{A}}

\newcommand{\Ann}{A}

\newcommand{\ann}{a^{{}}}

\newcommand{\BBB}{\myvec{B}}

\newcommand{\Bnn}{B}

\newcommand{\bnn}{b^{{}}}

\newcommand{\Cnn}{C^\notop}

\newcommand{\DDn}{\myvec{D}}

\newcommand{\EEn}{\myvec{E}}

\newcommand{\eee}{\myvec{e}}
\newcommand{\een}{\myvec{e}}

\newcommand{\FFFrad}{\myvec{F}^\mathrm{rad}}

\newcommand{\Frad}{F^{\mathrm{rad}}}

\newcommand{\GGn}{\myvec{G}}

\newcommand{\gnn}{g^{{}}}

\newcommand{\JJJ}{\myvec{J}}

\newcommand{\kc}{k_\mathrm{c}}

\newcommand{\ks}{k_\mathrm{s}}

\newcommand{\MMM}{\myvec{M}}

\newcommand{\nnn}{\myvec{n}}

\newcommand{\rrr}{\myvec{r}}
\newcommand{\rrrO}{\rrr_0}

\newcommand{\sss}{\myvec{s}}
\newcommand{\ssn}{\sss^\notop}
\newcommand{\ssspz}{\sss^{\mr{pz}_{}}}

\newcommand{\spz}{s^{\mr{pz}_{}}}

\newcommand{\ttt}{\myvec{t}}
\newcommand{\ttn}{\myvec{t}^\notop}

\newcommand{\uuu}{\myvec{u}}
\newcommand{\uun}{\myvec{u}^\notop}
\newcommand{\unn}{u^\notop}
\newcommand{\VVV}{\myvec{V}}

\newcommand{\WWW}{\myvec{W}}

\newcommand{\vvv}{\myvec{v}}
\newcommand{\vvn}{\myvec{v}^{{}}}
\newcommand{\vnn}{v^{{}}}

\newcommand{\vstr}{v_\mathrm{str}}

\newcommand{\zerovec}{\boldsymbol{0}}

\newcommand{\cO}{c_0}

\newcommand{\cfl}{c_\mr{fl}}

\newcommand{\Eac}{E_\mathrm{ac}}

\newcommand{\kapfl}{\kappa_\mr{fl}}

\newcommand{\pL}{p^{\longrange}}
\newcommand{\ptL}{\tilde{p}^{\longrange}}

\newcommand{\deltan}{\delta}

\newcommand{\eps}{\epsilon}

\newcommand{\epsO}{\epsilon_0}

\newcommand{\etaBO}{\eta^\mathrm{b}_0}

\newcommand{\etaO}{\eta_0}

\newcommand{\etafl}{\eta_\mr{fl}}
\newcommand{\etaflb}{\eta^{{\mr{b}}}_\mr{fl}}

\newcommand{\nuO}{\nu_0}

\newcommand{\Omegafl}{\Omega^\mr{fl}}
\newcommand{\Omegasl}{\Omega^\mr{sl}}
\newcommand{\Omegapz}{\Omega^\mr{pz}}

\newcommand{\vph}{\varphi}
\newcommand{\vphn}{\varphi^\notop}

\newcommand{\rhofl}{\rho^\mr{fl}}
\newcommand{\rhosl}{\rho^\mr{sl}}
\newcommand{\rhopz}{\rho^\mr{pz}}

\newcommand{\cOsqr}{c^{\,2_{}}_0}

\newcommand{\fO}{f_0}

\newcommand{\fI}{f_1}

\newcommand{\fres}{f_\mathrm{res}}

\newcommand{\kO}{k_0}

\newcommand{\kapO}{\kappa_0}

\newcommand{\pI}{p_1}

\newcommand{\pII}{p_2}

\newcommand{\vvvI}{\vvv^\notop_1}

\newcommand{\vvvII}{\vvv^\notop_2}

\newcommand{\rhoO}{\rho_0}

\newcommand{\rhoI}{\rho_1}
\newcommand{\rhoII}{\rho_2}

\newcommand{\SIC}{\textrm{C}}
\newcommand{\SICel}{^\circ\!\textrm{C}}
\newcommand{\SIF}{\textrm{F}}

\newcommand{\SIMHz}{\textrm{MHz}}

\newcommand{\SIJpcm}{\textrm{J}\:\textrm{m$^{-3}$}}

\newcommand{\SIkgm}{\textrm{kg}\:\textrm{m$^{-3}$}}
\newcommand{\SIkgpcm}{\SIkgm}

\newcommand{\SIm}{\textrm{m}}
\newcommand{\SIcm}{\textrm{cm}}
\newcommand{\SImm}{\textrm{mm}}
\newcommand{\SImum}{\textrm{\textmu{}m}}
\newcommand{\SInm}{\textrm{nm}}

\newcommand{\SIpN}{\textrm{pN}}

\newcommand{\SIkPa}{\textrm{kPa}}
\newcommand{\SIpTPa}{\textrm{TPa}^{-1}}
\newcommand{\SIMPa}{\textrm{MPa}}
\newcommand{\SIGPa}{\textrm{GPa}}
\newcommand{\SIPas}{\textrm{Pa}\:\textrm{s}}
\newcommand{\SImPas}{\textrm{mPa}\:\textrm{s}}

\newcommand{\SIs}{\textrm{s}}

\newcommand{\SImps}{\SIm\,\SIs^{-1}}
\newcommand{\SImumps}{\SImum\,\SIs^{-1}}

\newcommand{\SImS}{\textrm{mS}}
\newcommand{\SIV}{\textrm{V}}

\newcommand{\nn}{\nonumber}
\newcommand{\beq}[1]{\begin{equation} \eqlab{#1}}
\newcommand{\eeq}{\end{equation}}
\newcommand{\bsub}{\begin{subequations}}
\newcommand{\esub}{\end{subequations}}
\def\bal#1\eal{\begin{align}#1\end{align}}
\def\balat#1#2\ealat{\begin{alignat}{#1} #2 \end{alignat}}
\def\bsubal#1 #2\esubal{\bsuba{#1}\begin{align}#2\end{align} \esuba}     
\def\bsubalat#1#2#3\esubalat{\bsuba{#1} \begin{alignat}{#2} #3 \end{alignat} \esub}
\newcommand{\bsuba}[1]{\bsub \eqlab{#1}}
\newcommand{\esuba}{\esub}

\newcommand{\emiot}{\ee^{-\ii\omega t}}

\newcommand{\eqlab}[1]{\label{eq:#1}}
\renewcommand{\eqref}[1]{Eq.~(\ref{eq:#1})}
\newcommand{\eqnoref}[1]{(\ref{eq:#1})}

\newcommand{\eqsref}[2]{Eqs.~(\ref{eq:#1}) and~(\ref{eq:#2})}

\newcommand{\eqssref}[3]{Eqs.~(\ref{eq:#1}), (\ref{eq:#2}) and~(\ref{eq:#3})}

\newcommand{\figref}[1]{Fig.~\ref{fig:#1}}

\newcommand{\figsref}[2]{Figs.~\ref{fig:#1} and~\ref{fig:#2}}
\newcommand{\figlab}[1]{\label{fig:#1}}
\newcommand{\appref}[1]{Appendix~\ref{sec:#1}}

\newcommand{\appsssref}[4]{Appendices~\ref{sec:#1}, \ref{sec:#2}, \ref{sec:#3}, and~\ref{sec:#4}}

\newcommand{\secref}[1]{Sec.~\ref{sec:#1}}

\newcommand{\secsref}[2]{Secs.~\ref{sec:#1} and~\ref{sec:#2}}
\newcommand{\secXref}[2]{Secs.~\ref{sec:#1}-\ref{sec:#2}}

\newcommand{\seclab}[1]{\label{sec:#1}}
\newcommand{\tabref}[1]{Table~\ref{tab:#1}}

\newcommand{\tabsref}[2]{Tables~\ref{tab:#1} and~\ref{tab:#2}}
\newcommand{\tabssref}[3]{Tables~\ref{tab:#1}, \ref{tab:#2}, and~\ref{tab:#3}}
\newcommand{\tablab}[1]{\label{tab:#1}}

\newcommand{\sigmabf}{\bm{\sigma}}

\newcommand{\sigmafl}{\sigma^\fl}
\newcommand{\sigmasl}{\sigma^\sl}
\newcommand{\sigmapz}{\sigma^\pz}
\newcommand{\sigmabfsl}{\bm{\sigma}^\sl}
\newcommand{\sigmabffl}{\bm{\sigma}^\fl}
\newcommand{\sigmabfpz}{\bm{\sigma}^\pz}

\newcommand{\uuuI}{\myvec{u}_1}

\newcommand{\fl}{\mathrm{fl}}
\newcommand{\pz}{\mathrm{pz}}

\renewcommand{\sl}{\mathrm{sl}}
\newcommand{\wl}{\mathrm{wl}}
\newcommand{\gr}{\mathrm{gr}}
\newcommand{\AlScN}{\text{Al$_{0.6}$Sc$_{0.4}$N}}

\newcommand{\edb}[1]{{\color{blue} #1}}

\definecolor{darkgreen}{rgb}{0.00, 0.50, 0.00}
\definecolor{DARKGREEN}{rgb}{0.00, 0.50, 0.00}
\definecolor{RED}{rgb}{1.00, 0.00, 0.00}
\definecolor{GREEN}{rgb}{0.00, 1.00, 0.00}
\definecolor{BLUE}{rgb}{0.00, 0.00, 1.00}
\definecolor{MAGENTA}{rgb}{1.00, 0.00, 1.00}

\newcommand{\ord}[1]{\mathcal{O}({#1})}

\newcommand{\shortrange}{\delta} 			
\newcommand{\longrange}{d}				
\newcommand{\atsurface}{0}				

\newcommand{\vvvwall}{\VVV^\atsurface}				

\newcommand{\xs}{\xi}
\newcommand{\ys}{\eta}
\newcommand{\zs}{\zeta}
\newcommand{\AAAs}{\AAA^\atsurface}
\newcommand{\BBBs}{\BBB^\atsurface}
\newcommand{\AAApar}{\AAA_\parallel}

\newcommand{\AAAspar}{\AAA_\parallel^\atsurface}

\newcommand{\pardiv}{\nablabf_{\parallel}^\notop\!\cdot}

\renewcommand{\perp}{\zeta}

\newcommand{\Bs}{B^\atsurface}
\newcommand{\pargrad}{\grad_\parallel}
\newcommand{\vvvl}{\vvv^\longrange}
\newcommand{\vd}{{v}^\shortrange}
\newcommand{\vvvd}{\vvv^\shortrange}
\newcommand{\vvvds}{\vvv^{\shortrange\atsurface}}

\newcommand{\vvvdspar}{\vvv^{\shortrange\atsurface}_{1\parallel}}

\newcommand{\vvvddspar}{\vvv^{\shortrange\atsurface}_{2\parallel}}
\newcommand{\vvvls}{\vvv^{\longrange\atsurface}}
\newcommand{\vdsperp}{{v}^{\shortrange\atsurface}_{1\perp}}
\newcommand{\vlsperp}{{v}^{\longrange\atsurface}_{1\perp}}
\newcommand{\vddsperp}{{v}^{\shortrange\atsurface}_{2\perp}}

\newcommand{\vwallperp}{\vwall_{1\perp}}
\newcommand{\vvvwallpar}{\vvvwall_{1\parallel}}

\newcommand{\vvvlspar}{\vvv^{\longrange\atsurface}_{1\parallel}}

\newcommand{\vl}{{v}^\longrange}
\newcommand{\vwall}{V^\atsurface}					

\newcommand{\ezs}{\een_\zs}
\newcommand{\pd}{p^{\shortrange}}
\newcommand{\sigmabfd}{\sigmabf^\shortrange}
\newcommand{\sigmabfl}{\sigmabf^\longrange}
\newcommand{\ppperp}{\partial^\notop_\perp}
\newcommand{\ppperpsqr}{\partial_\perp^2}

\newcommand{\vvvddp}{\vvv_{2}^{\shortrange p}}
\newcommand{\vvvddv}{\vvv_{2}^{\shortrange v}}

\newcommand{\vvvddparp}{\vvv_{2\parallel}^{\shortrange p}}
\newcommand{\vvvddparv}{\vvv_{2\parallel}^{\shortrange v}}
\newcommand{\vvvddparpO}{\vvv_{2\parallel}^{\shortrange p 0}}
\newcommand{\vvvddparvO}{\vvv_{2\parallel}^{\shortrange v 0}}
\newcommand{\vddperp}{\vd_{2\perp}}
\newcommand{\vddperpO}{{v}_{2\perp}^{\shortrange 0}}
\newcommand{\vddperppO}{{v}_{2\perp}^{\shortrange p 0}}
\newcommand{\vddperpp}{{v}_{2\perp}^{\shortrange p}}
\newcommand{\vddperpv}{{v}_{2\perp}^{\shortrange v}}
\newcommand{\vddperpvO}{{v}_{2\perp}^{\shortrange v 0}}

\newcommand{\vdsCperp}{{v}^{\shortrange \atsurface*}_{1\perp}}

\newcommand{\vds}{{v}^{\shortrange\atsurface}}

\newcommand{\gd}{\gshort^{\shortrange}}

\newcommand{\gshort}{g}				

\begin{document}

\title{Boundary-layer modeling of polymer-based acoustofluidic devices}

\author{Sazid Z. Hoque}
\email{shoque@dtu.dk}
\affiliation{Department of Physics, Technical University of Denmark,\\
DTU Physics Building 309, DK-2800 Kongens Lyngby, Denmark}

\author{Henrik Bruus}
\email{bruus@fysik.dtu.dk}
\affiliation{Department of Physics, Technical University of Denmark,\\
DTU Physics Building 309, DK-2800 Kongens Lyngby, Denmark}

\date{7 July 2025, DRAFT}

\begin{abstract}
In fluid-filled microchannels embedded in solid devices and driven by MHz ultrasound transducers, the thickness of the viscous boundary layer in the fluid near the confining walls is typically 3 to 4 orders of magnitude smaller than the acoustic wavelength and 5 orders of magnitude smaller than the longest dimension of the device. This large span in length scale renders direct numerical simulations of such devices prohibitively expensive in terms of computer memory requirements, and consequently, the so-called boundary-layer models are introduced. In such models, approximate analytical expressions of the boundary-layer fields are found and inserted in the governing equations and boundary conditions for the remaining bulk fields. Since the bulk fields do not vary across the boundary layers, they can be computed numerically using the resulting boundary-layer model without resolving the boundary layers. However, current boundary-layer models are only accurate for hard solids (\eg\ glass and silicon) with relatively small oscillation amplitudes of the confining wall, and they fail for soft solids (\eg\ polymers) with larger wall oscillations. In this work, we extend the boundary-layer model of Bach and Bruus, J.~Acoust.\ Soc.\ Am.\ \textbf{144}, 766 (2018) to enable accurate simulation of soft-walled devices. The extended model is validated by comparing (1) with direct numerical simulations in three and two dimensions of tiny sub-mm and larger mm-sized polymer devices, respectively, and (2) with previously published experimental data.
\end{abstract}

\maketitle


%
%

\section{Introduction}
\seclab{intro}
Acoustofluidic devices based on bulk acoustic waves show promising applications in contactless manipulation of micro-objects \cite{Laurell2014, Fan2022} such as separation of bacteria from blood lysate \cite{VanAssche2020}, cancer cells enrichment in blood \cite{Anand2021, Anwar2024}, optimized microparticle focusing \cite{Qiu2022, Baasch2024}, and other cell manipulations \cite{Wu2019}. These devices are also used for manipulation of immiscible fluid interfaces \citep{Hemachandran2021, Hoque2024} and inhomogeneous fluids \cite{Deshmukh2014, Karlsen2016, Augustsson2016, Karlsen2018}.

Traditionally, bulk-acoustic-wave devices are fabricated using hard materials with high Q factors, such as glass and silicon \cite{Lenshof2012}. Although such devices can be manufactured with high accuracy and throughput, the fabrication process may be expensive, limiting in particular their single-use applications. Alternatively, polymer-based devices can be used, which is advantageous for bulk production at minimal cost compared to glass-silicon devices \cite{Savage2017, Yang2017, Silva2017, Dubay2019, Lickert2021}. However, it is difficult to establish the acoustic fields in a polymer system due to the low acoustic contrast between the polymer and the fluid. Recently, the formation of  pressure nodal planes in polymer systems has been elucidated by Moiseyenko and Bruus \cite{Moiseyenko2019} by introducing the principle of whole-system ultrasound resonances (WSUR). According to this principle, acoustic resonances in polymer devices are determined by the dimensions of the whole system, and by the acoustic contrast between the ambient air and the system. Among the obtained WSUR modes, the ones leading to a robust acoustic response inside the fluid cavity are selected. Successful focusing of particles in polymer-based acoustofluidics devices designed using the WSUR principle has been demonstrated by experiments and simulations \cite{Lickert2021}.

Despite the significant advancement in numerical modeling of acoustofluidics devices in the last decade \cite{Muller2012, Muller2013, Muller2014, Hahn2015, Ley2016}, it remains a challenge to have a robust, computationally efficient method that captures all the relevant length scales of the problem ranging from the length of the device ($\sim 5~\SIcm$) down to the thickness of the viscous boundary layers ($\sim 0.5~\SImum$) in the fluid, the latter being responsible for the formation of the steady acoustic streaming. Whereas the pressure acoustics, and thus the acoustic radiation force that dominates the dynamics of large particles ($\gtrsim 2~\SImum$) suspended in the fluid, depends on the acoustic wavelength ($\sim 500~\SImum$) and not on the acoustic streaming, the dynamics of sub-$\SImum$ particles are dominated by the viscous Stokes drag from the acoustic streaming. The inclusion of the boundary layers in numerical simulations is therefore crucial for a full understanding of acoustofluidic devices and their applications for sub-$\SImum$-particle manipulation in lab-on-a-chip technology \cite{Hammarstrom2012, Antfolk2015, Mao2017, Collins2017}. Performing direct numerical simulations to capture the entire length-scale range is computationally expensive and poses severe challenges, especially for three-dimensional (3D) simulations.

The effective boundary-layer model introduced by Bach and Bruus in 2018 \cite{Bach2018} marked a significant progress towards robust modeling of the thin boundary layer. Based on previous work by Nyborg \cite{Nyborg1958}, Lee and Wang \cite{Lee1989}, and Vanneste and  B\"{u}hler \cite{Vanneste2011}, Bach and Bruus derived approximate analytical expressions for the viscous boundary-layer fields that decays exponentially away from the fluid-solid interface inside the thin boundary layer. Inserting these analytic expressions in the governing equations and boundary conditions, an effective model for the remaining bulk acoustic fields were obtained. As these bulk fields do not vary on the small boundary-layer length scale, they can be computed numerically without resolving the boundary layer.

The effective boundary-layer model is based on the assumptions that the boundary-layer thickness $\delta$ is much smaller than both the radius of curvature of the fluid-solid interface (the wall) and the acoustic wavelength, and that the amplitude of the wall motion in the perpendicular direction is much smaller than $\delta$. 3D versions of the model have been validated against experimental results in two cases. First, Skov \etal~\cite{Skov2019} used the model to simulate a hard-walled glass-silicon device and successfully compared their results with previous experimental work \cite{Hagsater2007}, both regarding the 1st-order acoustic fields and the 2nd-order streaming field, although the latter is sensitive to the exact value of the resonance frequency and of detailed shape of the fluid-solid interface. Second, Lickert \etal~\cite{Lickert2021} compared simulation results of the model for a soft-walled polymer-based device, and found fairly good agreement with their own experiments regarding the 1st-order acoustic fields, but no studies of the streaming field was performed.

The main goal of this work is to evaluate the 2nd-order time-averaged streaming field using the boundary-layer model for a soft-walled polymer device. We have found that the second assumption above concerning the smallness of the wall oscillation amplitude compared to $\delta$ fails in this case, and therefore an extension of the boundary-layer model is needed to accurately determining the acoustic streaming field for soft-walled devices. In this work, we extend the boundary-layer model from 2018 by Bach and Bruus \cite{Bach2018}, henceforth called the \qm{BL18 model}, by allowing for larger perpendicular wall motions. Introduced in 2025, we call this extended boundary-layer model the \qm{BL25 model}.

The first half of the paper, \secXref{pert_theory}{fl_second_order}, concerns the development of the theory behind the BL25 model.
The basic acoustofluidic perturbation approach as well as the detailed 1st-order expressions for the acoustic fields  carry over unchanged (with one minor addition in a boundary condition) from the BL18 model to the BL25 model, which is summarized in \secsref{pert_theory}{fl_first_order}. In \secref{fl_second_order}, we present the model of the extended 2nd-order streaming field, where we modify the effective slip-boundary condition by (1) taking into account the hitherto neglected short-range 2nd-order pressure field, and (2) including previously omitted higher-order terms in the small parameter $\epsilon = \kO\delta \simeq 0.003$, $\kO$ being the acoustic wavenumber. The second half of the paper, \secXref{device_model}{expValidation}, concerns the validation of the BL25 model. In \secsref{device_model}{COMSOL_impl}, respectively, we define the acoustofluidic model devices used for numerical validation of the BL25 model, and we present the implementation of the model for these devices in the finite-element software COMSOL Multiphysics. In \secref{numValidation}, we carry out a numerical validations including a mesh-convergence study of the BL25 model in both two-dimensional (2D) and 3D simulations by comparing it to a \qm{Full model} consisting of direct numerical simulation of the full perturbation equations using a mesh that resolves the thin boundary layers. The 3D simulation of the Full model is made possible by studying a down-scaled device with a linear size less then $500~\SImum$, which is driven by a thin-film piezoelectric transducer of thickness $2~\SImum$. In \secref{expValidation}, we validate the BL25 model experimentally by using it to model the polymer-device studied by Lickert \etal~\cite{Lickert2021} and comparing the simulation results with the published experimental data. Finally, we end with a summary and concluding remarks in \secref{conclusion}. Some mathematical details are presented in \appsssref{BC2ndOrderAppendix}{constitutiveEqs}{COMSOLimplem}{MeshConvergence}, and supplementing numerical simulations are provided in the Supplemental Material~\footnote{See Supplemental Material at
\url{https://bruus-lab.dk/Hoque-BL25-Suppl.pdf} 
for an analysis of corner effects in rectangular channels and for details on the dependency on the glue-layer thickness.}, which includes Refs.~\cite{CorningPyrex, Lickert2021}.

%
%

\section{Perturbation expansion of the governing equations}
\seclab{pert_theory}

In this work, we follow Bach and Bruus \cite{Bach2018} closely, and present here a just summary of the BL18 model, while referring the reader to the original paper for details. We consider a system consisting of a fluid domain $\Omegafl$ embedded in an elastic solid domain $\Omegasl$, which is actuated by a piezoelectric transducer domain $\Omegapz$ attached to a part $\pp \Omegasl_\mr{pz}$ of the outer solid surface $\pp\Omegasl$ and driven by an applied time-harmonic AC voltage $\vphn_0(t) = \vphn_0\:\emiot$ with a real-valued amplitude $\vphn_0$ at the angular frequency $\omega = 2\pi f$, where $f$ is the frequency in the MHz range. The fluid is described by three basic Eulerian fields, the density $\rho$, the pressure $p$, and the velocity $\vvv$, and the solid by the basic Lagrangian field, the displacement $\uuu$. We assume adiabatic acoustics without thermal transport, so the temperature field is not independent, but follows from the pressure and density. Assuming a sufficiently low actuation amplitude $\vphn_0$, the whole system can be analyzed using a complex-valued perturbation expansion, where any field $g$ depending on position $\rrr$ and time $t$ has a uniform constant 0th-order component $g_0$, a time-harmonic 1st-order component $g_1$, and a steady time-averaged 2nd-order component $g_2$, proportional to $(\vphn_0)^0$, $(\vphn_0)^1$, and $(\vphn_0)^2$, respectively,
 \bsubal{PertExpansion}
 \eqlab{gExpanded}
 g(\rrr,t) &= g_0 + g_1(\rrr)\:\emiot 	+ g_2(\rrr),
 \\
 \eqlab{gPhysExpanded}
 g^\mr{phys}(\rrr,t)  &= g_0 + \re\big[g_1(\rrr)\:\emiot\big] 	+ g_2(\rrr).
 \esubal
Here, the physical field $g^\mr{phys}(\rrr,t)$ is simply obtained as the real part of the complex-valued field $g(\rrr,t)$. Note that both the 0th-order term $g_0(\rrr)$ and the time-averaged 2nd-order terms $g_2(\rrr)$ are real-valued and time independent. The time average $\avr{\cdots}$ is defined over a full oscillation period $\tau_0 = 2\pi/\omega$, and the time average of a product of two 1st-order fields is easily computed using the complex-valued fields,
 \bsubal{timeaverage}
 \eqlab{timeavrDef}
 \avr{g^\mr{phys}(\rrr,t)} &= \frac{1}{\tau_0}\int_0^{\tau_0} g^\mr{phys}(\rrr,t)\: \dm t,
 \\
 \eqlab{avrA1B1}
 \avr{A_1^\mr{phys}(\rrr,t)B_1^\mr{phys}(\rrr,t)} &= \frac12 \re\big[A^{{}}_1(\rrr)B^*_1(\rrr)\big],
 \esubal
where the asterisk denotes complex conjugation.

\subsection{The linear isotropic elastic solid}
\seclab{solidDomain}

We assume that the solid domain $\Omegasl$ contains a linear elastic solid with density $\rhosl$, where the position $\rrr$ of a solid element with equilibrium postion $\rrrO$ is given in terms of the Lagrangian displacement field $\uuu = \uuuI$ as
 \beq{uIdef}
 \rrr(\rrrO) = \rrrO + \uuuI(\rrrO)\:\emiot,\quad \text{for } \rrrO \in \Omegasl.
 \eeq
This is a 1st-order expression, since the steady time-averaged 2nd-order displacement $\uuu_2$ is neglected as the steady pressure $\pII$ in the fluid is too small to deform the surrounding solid noticeably. Using index notation, the governing equation for $\uuuI$ is given by the Cauchy momentum equation in terms of the elastic moduli $\Cnn_{ik}$ and the stress tensor $\sigmabf^\sl_1$ with components $\sigmasl_{1ik}$ \cite{Bode2022},
 \bsubal{CauchyEqu}
 \unn_{1i} &= -\frac{1}{\omega^2\rhosl}\:\pp_k\sigmasl_{1ik},\quad \text{for } \rrrO \in \Omegasl,
 \\
 \sigmasl_{1ik} &= \Cnn_{44}(\pp_i \unn_{1k} \!\!+\! \pp_k \unn_{1i})
 + (\Cnn_{11}\!\!-\!2\Cnn_{44})\pp_j \unn_{1j}\deltan_{ik}.
 \esubal

At the fluid-solid interface $\pp\Omegafl_\mr{sl}$, the equilibrium position is denoted $\ssn_0$, and the interface displacement and velocity is $\ssn_1 = \uuuI(\ssn_0)$ and $\vvvwall_1 = \pp_t\ssn_1$, respectively,
 \bsubal{sssV0}
 \eqlab{sss_eq} 
 \sss(\ssn_0,t) &= \ssn_0 + \ssn_1(\ssn_0)\:\emiot,\quad \text{for } \ssn_0 \in \pp\Omegafl_\mr{sl},
 \\
 \eqlab{dtsss} 
 \ppt \sss(\sss_0,t) &=-\ii\omega \sss_1(\sss_0)\: \emiot =\vvvwall_1(\sss_0)\:\emiot.
 \esubal
The boundary conditions on $\pp\Omegafl_\mr{sl}$ follows from \eqref{dtsss}, and they are treated further in \secref{fluidDomain}. The remaining part of $\pp\Omegasl$ is divided into the free surface $\Omegasl_\mr{free}$ and the transducer-solid interface $\Omegapz_\sl$,
 \bsubalat{BCsolid}{3}
 \eqlab{BCslfl}
 \sigmabfsl_1\cdot\nnn &= \sigmabffl_1\cdot\nnn \;\text{ and }\; \vvvwall_1 &&= \vvvI, \;
 &&\text{for } \rrr \in \pp\Omegafl_\sl,
 \\
 \eqlab{BCslpz}
 \sigmabfsl_1\cdot\nnn &= \sigmabfpz_1\cdot\nnn \;\text{ and }\; \uuu^\sl_1 &&= \uuu^\pz_1, \;
 &&\text{for } \rrr \in \pp\Omegapz_\sl,
 \\
 \eqlab{BCslfree}
 \sigmabfsl_1\cdot\nnn &= \zerovec, &&
 &&\text{for } \rrr \in \pp\Omegasl_\mr{free}.
 \esubalat

\subsection{The linear piezoelectric transducer}
\seclab{piezoDomain}

The linear piezoelectric transducer is described by the elastic deformation $\uun_1$ and the electric potential $\vphn_1$ governed by the Cauchy momentum equation and the zero-free-charge Gauss's law for dielectrics coupled by the constitutive equations $\sigmabfpz_1(\uun_1,\vphn_1)$ and $\DDn_1(\uun_1,\vphn_1)$ for the mechanical stress and electric displacement, respectively, presented in \appref{constitutiveEqs},
 \bsubal{transducerEq}
 -\rhopz \omega^2\: \uun_1 &= \div\sigmabfpz_1(\uun_1,\vphn_1),
 \\
 \zerovec &= \div\DDn_1(\uun_1,\vphn_1),
 \esubal
where $\sigmabfpz_1(\uun_1,\vphn_1)$ and $\DDn_1(\uun_1,\vphn_1)$ are given in \eqref{StressStrainPiezo}.

The boundary $\pp\Omegapz$ of $\Omegapz$ is divided into three parts: the interface $\Omegapz_\sl$ between $\Omegapz$ and $\Omegasl$, the interface $\Omegapz_{\mr{el},n}$ between $\Omegapz$ and the $n$ solid electrodes, and the free surface $\Omegapz_\mr{free}$. The boundary conditions on each part are
 \bsubalat{BCpiezo}{2}
 \eqlab{BCpzElec} \nn
 \sigmabfpz_1\cdot\nnn &= \sigmabfsl_1\cdot\nnn,\; \uuu^\pz_1 = \uuu^\sl_1, \;
 &&\text{and } \vphn_1 = \vph^{(n)}_0,
 \\
 & &&\text{for } \rrr \in \pp\Omegapz_{\mr{el},n},
 \\ \nn
 \sigmabfpz_1\cdot\nnn &= \sigmabfsl_1\cdot\nnn,\; \uuu^\pz_1 = \uuu^\sl_1, \;
 &&\text{and } \nnn\cdot\DDn_1 = 0,
 \\
 & &&\text{for } \rrr \in \pp\Omegapz_\sl,
 \\
 \sigmabfpz_1\cdot\nnn &= \zerovec  \text{ and } \nnn\cdot\DDn_1 = 0,\;
 &&\text{for } \rrr \in \pp\Omegapz_\mr{free}.
 \esubalat

\subsection{The Newtonian fluid}
\seclab{fluidDomain}

We assume that the fluid domain $\Omegafl$ contains a Newtonian fluid described by the Eulerian fields density $\rho$, pressure $p$, and velocity $\vvv$, and by the parameters of the unperturbed state: density $\rhoO$, sound speed $\cO$, compressibility $\kapO = 1/(\rhoO\cOsqr)$, dynamic viscosity $\etaO$, and bulk viscosity $\etaBO$. The governing equations in the adiabatic limit are the continuity and momentum equation,
 \bsuba{GovEqu}
 \eqlab{cont_navier} 
 \bal
 \eqlab{cont} 
 \ppt\rho	&=-\div(\rho\vvv),
 \\
 \eqlab{navier} 
 \ppt(\rho\vvv)&= -\div[(\rho \vvv)\vvv] + \div\sigmabffl,
 \eal
where $\sigmabffl$ is the stress tensor with components $\sigmafl_{ik}$,
 \bal
 \eqlab{sigma} 
 \sigmafl_{ik} = \big[-p+\!(\etaBO\!-\!\tfrac23\etaO)\pp_j\vnn_j\!\big]\deltan_{ik}
 + \etaO (\pp_i\vnn_k\!+\!\pp_k\vnn_i).
 \eal
 \esuba
In contrast to the solid domain, the governing equations in the fluid domain are nonlinear, and thus we include the 2nd-order time-averaged fields in the perturbation expansion,
 \bsubalat{def_pert_series}{3} 
 \eqlab{def_rho} 
 \rho  &= \rho_0	&& + \rhoI(\rrr)\:\emiot && + \rhoII(\rrr),
 \\
 \eqlab{def_p} 
 p &= p_0 &&+ \pI(\rrr)\:\emiot && + p_2(\rrr),
 \\
 \eqlab{def_v} 
 \vvv &= \zerovec 	&&+ \vvn_1(\rrr)\:\emiot &&+ \vvn_2(\rrr),
 \quad \text{for  } \rrr \in \Omegafl.
 \esubalat

The resulting governing equations for 1st- and 2nd-order perturbation are presented in respectively \secsref{fl_first_order}{fl_second_order} below, but already here, we follow up on \eqref{sssV0} and formulate the no-slip boundary condition (no-slip BC) $\vvv = \pp_t\sss$  relating the Eulerian fluid velocity $\vvv$ to the Lagrangian wall velocity $\pp_t\sss$ \eqref{dtsss} at the fluid-solid interface $\pp\Omegafl_\mr{sl}$. This condition applies at all times at the actual position $\sss(\ssn_0,t)$ \eqref{sss_eq}, so we obtain
 \bal\eqlab{no-slip} 
 \vvv(\sss_0+\sss_1\emiot,t)=\vvvwall_1(\sss_0)\:\emiot,\: \text{no-slip BC}.
 \eal
Combining \eqsref{def_v}{no-slip} with the Taylor expansion
$\vvv(\sss_0+\sss_1,t) \approx \vvv_1(\sss_0)\:\emiot + \avr{(\sss_1\scap\grad)\vvv_1}\big|^{{}}_{\sss_0}$, and collecting the terms order by order, gives
 \bsuba{bc} 
 \balat{2}{
 \eqlab{bc_v1} 
 \vvv_1(\sss_0) &= \vvvwall_1(\sss_0),
 && \text{ 1st-order no-slip BC},
 \\
 \eqlab{bc_v2} 
 \vvv_2(\sss_0) &= -\avr{(\sss_1\scap\grad)\vvv_1}\big|^{{}}_{\sss_0},
 && \text{ 2nd-order no-slip BC}.
 }\ealat
 \esuba

\subsection{Surface, boundary-layer, and bulk fields for weakly-curved fluid-solid interfaces}
\seclab{WeaklyCurvedInterfaces}

A central quantity in acoustofluidic theory, is the so-called boundary-layer thickness $\delta$, which is a dynamically defined length scale that appears in the acoustically oscillating fluid inside $\Omegafl$ near the fluid-solid interface $\pp\Omegafl_\mr{sl}$, see \secref{fl_first_order}. For MHz-ultrasound in water, $\delta \approx 0.5~\SImum$ is very small compared to the inverse wave number $\kO^{-1} = \cO/\omega \approx 240~\SImum$, and for weakly curved surfaces it is also much smaller than the radius of curvature $R$ of $\pp\Omegafl_\mr{sl}$,
 \bsubalat{deltadeps}{2}
 \eqlab{deltaDef} 
 \delta & = \sqrt{\frac{2\nuO}{\omega}},\;
 && \text{small boundary length scale},
 \\
 \eqlab{dDef}
 d & = \min\!\Big\{\frac{\cO}{\omega}, R \Big\},\;\;
 && \text{large bulk length scale},
 \\
 \eqlab{epsdef} 
 \eps & = \frac{\delta}{d}\ll 1,
 && \text{small length-scale ratio}.
 \esubalat

As in BL18, we introduce for a given point $\ssn_0$ at $\pp\Omegafl_\mr{sl}$  the orthonormal local coordinates $\xs$, $\ys$, and $\zs$ with their origin at $\ssn_0$, such that  $\zeta$ is the normal coordinate pointing into $\Omegafl$, $\een_\zeta = -\nnn(\pp\Omegafl_\mr{sl})$, and $\xi$ and $\eta$ are the coordinates in the tangent plane to $\pp\Omegafl_\mr{sl}$ at $\ssn_0$ with $\een_\xi \times \een_\eta = \een_\zeta$.

Importantly, for fluid fields both in 1st- and in 2nd-order perturbation, we distinguish between bulk fields $\AAA^d$ that extend into the bulk with spatial variation on the bulk length scale $d$ and that are typically found by numerical simulation, and boundary-layer fields $\AAA^\delta$ that decays to zero away from the wall on the boundary-layer length scale $\delta$. For the boundary-layer fields, the smallness of the length ratio $\eps \sim \delta/d$ allows for separation of variables and simplification of the differential geometrical properties, as summarized in the following expressions, where superscript '0' refers to surface properties that only depends on the in-plane coordinates $\xs$ and $\ys$, and where subscript "$\parallel$" denotes tangential components,
 \bsuba{ABoundarySurface}
 \bal
 \eqlab{AdeltaDef} 
 \AAA^\delta(\xs,\ys,\zs) &\approx \AAA^{\delta 0} (\xs,\ys)\: a^\delta(\zs),
 \\ \nn
 & \qquad \text{with } a^\delta(\zs) \rightarrow 0 \text{ for } \frac{\zeta}{\delta} \rightarrow \infty.
 \\
 \eqlab{AAApar} 
 \AAApar &= A_\xs\:\een_\xs+A_\ys\:\een_\ys,
 \\
 \eqlab{nablapar} 
 \nablabf_\parallel &= \een_\xs \pp_\xs + \een_\ys \pp_\ys,
 \\
 \eqlab{curvi_derivatives_zero_a} 
 \div \AAAs &\approx \pardiv \AAAspar,
 \\
 \eqlab{curvi_derivatives_zero_b} 
 \AAAs \scap \grad \BBBs  &\approx \AAAspar \scap (\pargrad \Bs_i)\:\eee_i,
 \\
 \eqlab{curvi_lap_delta} 
 \Lapl a^\delta &\approx \ppperpsqr a^\delta,
 \\
 \eqlab{curvi_lap_delta_vector} 
 \Lapl \AAA^\delta & \approx \AAA^{\delta 0}\ppsqr_\zs a(\zs)=\ppsqr_\zs \AAA^\delta,
 \\
 \eqlab{curvi_div_delta} 
 \div\AAA^\shortrange &\approx \pardiv \AAA_\parallel^\shortrange
 +\ppperp A_\perp^\shortrange.
 \eal
 \esuba

%
%

\section{1st-order acoustic fluid fields}
\seclab{fl_first_order}

The first step in developing the governing equations~\eqnoref{GovEqu} for a Newtonian fluid in a 1st-order time-harmonic perturbation expansion~\eqnoref{def_pert_series}, is to use the adiabatic relation between the pressure $\pI$ and the density $\rhoI$,
 \bsubal{cont_navier_1}
 \eqlab{cont_1} 
 \ii\omega\kapO\pI &= \div\vvvI,\; \text{ with }\; \pI = \cOsqr \rhoI,
 \\
 \eqlab{navier_1}
 -\ii \omega \rhoO\vvvI &= \div\sigmabffl_1,
 \esubal
where the 1st-order stress $\sigmabffl_1$ is given by \eqref{sigma} upon substituting $p$ and $\vvv$ by $\pI$ and $\vvvI$, respectively. We have no changes to the treatment in the BL18 model of the 1st-order acoustic fields, so this section contains just a brief summary. The second step is to perform a Helmholtz decomposition of the velocity field $\vvvI$ into a curl-free compressible potential-flow component $\vvvl_1$ and a divergence-free incompressible component $\vvvd_1$,
 \bal
 \eqlab{Helmholtz_decomp} 
 \vvv_1 &= \vvvl_1+\vvvd_1,
 \\ \nn
 &\quad\;\text{with $\curl \vvvl_1=\zerovec$ and $\div\vvvd_1 =0$}.
 \eal
Using these expressions, $\vvvd_1$ decouples from $\vvvl_1$, $\pI$, and $\rhoI$, and drops out of the continuity equation, $\vvvl_1$ is proportional to $\nablabf\pI$, and the governing equations for $\pI$ and $\vvvd_1$ reduce to Helmholtz equations with the weakly damped compression wave number $\kc$ and the strongly damped shear wave number $\ks$, respectively,
 \bsubal{First-order_final} 
 \eqlab{v1dgradp1} 
 \vvvl_1 &=  -\ii\frac{1-\ii\Gamma}{\omega\rho_0}\:\grad \pI,
 \:\text{with } \Gamma = \frac{(1\!+\!\beta)\etaO\omega}{\rhoO\cOsqr},
 \\
 \eqlab{p1divvld} 
 \div \vvvl_1 &= \ii\omega\kapO\: p_1,
 \:\text{with } \beta = \frac{\etaBO}{\etaO}\!+\!\frac13,
 \\
 \eqlab{Helmholtz_p1} 
 \Lapl p_1 &= -\kc^2 p_1,
 \text{with }  \kc = \Big(1+\ii\frac{\Gamma}{2}\Big) k_0,\: \kO = \frac{\omega}{\cO},
 \\
 \eqlab{Helmholtz_vd} 
 \Lapl \vvvd_1 &= -\ks^2\vvvd_1,
 \text{with } \ks =\frac{1+\ii}{\delta},\: \delta = \sqrt{\frac{2\nu_0}{\omega}}.
 \esubal

We note that it is through the Helmholtz equation~\eqnoref{Helmholtz_vd} with its exponentially decaying solutions due to the shear wave number $\ks$ that the notion of boundary layers of width $\delta$ enters the theory. We also note that the damping coefficient $\Gamma$ in the compression wave number $\kc$ is of second order in the small parameter~$\eps$ of \eqref{epsdef}, $\Gamma = \frac12 (1+\beta)\: \eps^2\ll 1$, indeed a weak damping.

Although of second order, the average acoustic energy density $\Eac$ of the fluid occupying the volume $V_\mr{fl}$ and the time-averaged acoustic radiation force $\FFFrad$ on a suspended particle of radius $a$ and scattering coefficients $\fO$ and $\fI$, are given directly by the 1st-order fields as~\cite{Lickert2021},
 \bsubal{EacFrad}
 \eqlab{EacDef}
 \Eac &=
 \frac{1}{V_\mr{fl}}\:\int_{V_\mr{fl}} \Big[\frac14\kapO |\pI|^2 + \frac14\rhoO|\vvvI|^2 \Big]\:\dm V,
 \\
 \eqlab{FradDef}
 \FFFrad &= \pi a^3\grad\Big[\frac13 \fO \kapO |\pI|^2  - \frac12 \fI \frac14\rhoO|\vvvI|^2 \Big].
 \esubal

\subsection{The analytical 1st-order boundary-layer field}
For weakly curved, thin boundary layers, discussed in \secref{WeaklyCurvedInterfaces}, the analytical form of the  incompressible velocity field $\vvvd_1$ is easily found from \eqref{Helmholtz_vd} to be
 \bsubal{vd_sol_combined} 
 \eqlab{vd_sol} 
 \vvvd_1 & = \vvvds_1(\xs,\ys)\: \ee^{\ii\ks\zs} + \ord{\eps},
 \\
 \eqlab{vvvds_sol} 
 \vvvds_1 &= \vvvwall_1 - \vvvls_1, \text{ 1st-order no-slip BC}.
 \esubal
Here, by combining \eqsref{bc}{Helmholtz_decomp}, the amplitude $\vvvds_1(\xi,\eta)$ of $\vvvd_1$ on the surface $\zs = 0$ is expressed as the difference between the oscillation velocity $\vvvwall_1(\xi,\eta)$ of the wall and the value $\vvvls_1(\xi,\eta)$ of the compressible (acoustic) velocity field $\vvvl_1$ at the surface. Note how $\vvvd_1 = \vvvds_1\:\ee^{\ii \zs/\delta}\:\ee^{-\zs/\delta} $ decays exponentially on the small length scale $\delta$ in the $\zs$-direction normal to the wall.

\subsection{Boundary condition for the 1st-order pressure}
\seclab{BC_p1}
A crucial relation in the BL18 model, is the expression for the normal component $\vdsperp$ of $\vvvds_1$ on the surface in terms of the wall velocity $\vvvwall$ and the bulk acoustic velocity $\vvvls_1$, which follows from the incompressibility condition~\eqnoref{Helmholtz_decomp} and the analytical form~\eqnoref{vd_sol_combined},
 \beq{vdsperp} 
 \vdsperp = \frac{\ii}{\ks}\pardiv \vvvdspar =\frac{\ii}{\ks}\pardiv \vvvwallpar-\frac{\ii}{\ks}\pardiv \vvvlspar.
 \eeq
Using this result, the boundary condition $\vlsperp = \vwallperp-\vdsperp$ for the bulk field at the surface takes the self-consistent form to leading order in $\delta = (1+\ii)/\ks$.
 \beq{vlsperp} 
 \vlsperp = \Big(\vwallperp-\frac{\ii}{\ks}\pardiv \vvvwallpar\Big)+\frac{\ii}{\ks}\pardiv \vvvlspar.
 \eeq
Writing this 1st-order velocity boundary condition (BC) in terms of the pressure $\pI$, using $\pardiv \vvvlspar=\div \vvvls_1-\ppperp \vlsperp$ and \eqref{v1dgradp1}, leads to the self-consistent boundary condition for the 1st-order pressure $\pI$ in the BL18 model,
 \bal\eqlab{p1_bc} 
 \ppperp{p_1} =&\; \frac{\ii\omega\rho_0}{1-\ii\Gamma} \Big(\vwallperp-\frac{\ii}{\ks}\pardiv \vvvwallpar \Big)-\frac{\ii}{\ks}\Big(\kc^2p_1+\ppperpsqr p_1\Big),\nn
 \\
 &\; \text{1st-order pressure BC at $\Omegafl$}.
 \eal

\subsection{Boundary condition for the 1st-order stress}
\seclab{BC_sigma1}
The 1st-order stress condition $\sigmabf^\sl\cdot\ezs= \sigmabffl_1\cdot\ezs$ at the interface $\pp\Omegafl_\mr{sl}$ with inward surface normal $\een_\zs$, is rewritten by combining the 1st-order part $\sigmabffl_1$ of $\sigmabffl$ \eqref{sigma} with the 1st-order expressions \eqssref{Helmholtz_decomp}{First-order_final}{vd_sol_combined}. Keeping terms to leading order in $\kO\delta$ leads to the 1st-order stress boundary condition of the BL18 model,
 \bal\eqlab{stress1_bc} 
 \sigmabf^\sl_1\cdot\ezs = -p_1\ezs\; +&\;
 \ii\ks \eta_0 \Big(\vvvwallpar-\vvvl_{1\parallel}\Big)
 \nn \\
 +&\;
 \ii \ks \eta_0 2\Big(\vwallperp-\vl_{1\zs}\Big)\:\een_\zs,
 \nn \\
  = -p_1\ezs\; +&\;
 \ii\ks \eta_0 \Big[\vvvwall-\vvvl_1 + \big(\vwallperp-\vl_{1\zs}\big)\een_\zs\Big],
 \nn \\
 &\;\text{1st-order stress BC at $\Omegafl$},
 \eal
where in the last expression have used $\vvv = \vvv_\parallel + v^{{}}_\zeta\een_\zs$.
Since $\ks\etaO = \frac{1+\ii}{\delta}\etaO = \frac{1+\ii}{2}\kO\delta\:\rhoO\cO$, \eqref{stress1_bc} is the usual pressure condition plus a correction term of order $\kO\delta = \eps$ due to the viscous stress $\etaO\ppperp\vvvd_{1}$ in the boundary layer. The last term in \eqref{stress1_bc} was left out in the BL18 model.

%
%

\section{2nd-order time-averaged streaming fields in the fluid}
\seclab{fl_second_order}
We compute the time-averaged acoustic streaming by 2nd-order perturbation theory, closely following the BL18 model
but now extending it in the BL25 model by keeping terms of higher-order in $\kO\delta$ hitherto neglected. The time-averaged part of the governing equations and the boundary conditions in 2nd-order perturbation are obtained from \eqsref{cont_navier}{bc},
 \bsubal{cont_navier_second} 
 \eqlab{CE_long} 
 \rho_0\div\vvv_2 &= -\div \timeav{\rho_1}{\vvv_1},
 \\
 \eqlab{NS_long} 
 \div\sigmabffl_2 &= \rho_0\div\timeav{\vvv_1}{\vvv_1},
 \\
 \eqlab{Langrage_bc_v2} 
 \vvv_2^0 &= - \frac{1}{\omega} \timeav{\ii\vvv_1^0 \cdot\grad}{\vvv^0_1},
 \esubal
where we have followed \eqref{def_pert_series} and dropped $\avr{\cdot}$ from the time-averaged velocity $\vvv_2$, pressure $\pII$, and stress $\sigmabffl_2$. Equation~\eqnoref{cont_navier_second} is a compressible Stokes flow problem in the 2nd-order fields $\vvvII$ and $\sigmabffl_2$, where the known 1st-order fields act as source terms. This problem is split in two parts: the 2nd-order boundary-layer fields driven by the short-range source terms $\divop\avr{\rhoI\vvvd_1}$ and $\rhoO\divop \avr{\vvvd_1\vvvd_1 + \vvvd_1\vvvl_1 + \vvvl_1\vvvd_1}$, and the 2nd-order bulk streaming field driven by the long-range source terms, $\divop\avr{\rhoI\vvvl_1}$ and $\rhoO\divop\avr{\vvvl_1\vvvl_1}$. The corresponding responses are long-range bulk fields (superscript \qm{$d$}) and  short-range boundary-layer fields (superscript \qm{$\delta$}),
 \bsubalat{decomp_second}{2} 
 \vvv_2&=\vvvl_2+\vvvd_2, && \text{for } \rrr \in \pp\Omegafl_\mr{sl},
 \\
 p_2 &= \pL_2+\pd_2, && \text{for } \rrr \in \pp\Omegafl_\mr{sl},
 \\
 \sigmabffl_2&=\sigmabfl_2+\sigmabfd_2, && \text{for } \rrr \in \pp\Omegafl_\mr{sl},
 \\
 \eqlab{decomp_bc} 
 \vvvls_2 &= -\vvvds_2 - \avr{(\sss_1\cdot\grad)\vvv_1}, \;\;
 && \text{for } \rrr = \sss_0.
 \esubalat

\subsection{Short-range boundary-layer streaming}
The governing equations for the 2nd-order boundary-layer fields, $\vvvd_2$, $\pd_2$, and $\sigmabfd_2$, are obtained by collecting the terms of \eqref{cont_navier_second} which contains at least one short-range 1st-order field  $\vvvd_1$,
 \bsuba{equMotion2}
 \eqlab{cont_navier_second_short} 
 \bal
 \eqlab{cont_navier_second_short_a} 
\div(\rho_0\vvvd_2) =&\; -\div\avr{\rho_1\vvvd_1},
 \\
 \eqlab{cont_navier_second_short_b} 
 \div\sigmabfd_2 =&\; \rho_0\div\avr{\vvvd_1\vvvd_1 + \vvvd_1\vvvl_1 + \vvvl_1\vvvd_1},
 \\
 \text{with } \eqlab{sigmad2} 
 \div \sigmabfd_2 =&\; \nablabf\big(-\pd_2+\beta\eta_0\div\vvvd_2\big)
 + \etaO \Lapl \vvvd_2,
 \\
 \eqlab{cont_navier_second_short_c} 
 \text{where } \vvvd_2\rightarrow \zerovec & \text{ and $\pd_2 \rightarrow 0$ as $\zs \rightarrow \infty$}.
 \eal
 \esuba
Thus $\vvvd_2$, $\pd_2$, and $\sigmabfd_2$ are driven by source terms containing time-average products of 1st-order terms of the form $\timeav{\AAA_1}{\BBB_1}$, where $\AAA_1$ and $\BBB_1$ are either $\vvvl_1$ or $\vvvd_1$. To solve \eqref{equMotion2}, we first combine \eqref{cont_navier_second_short_b} and \eqref{sigmad2} to obtain the Laplacian of $\vvvd_2$,
 \bal 
 \eqlab{Laplvd2}
 \etaO \Lapl \vvvd_2&= -\beta \etaO \grad(\div \vvvd_2)+ \grad \pd_2
 \nn \\
 & \quad + \rhoO \div \avr{\vvvd_1 \vvvd_1+ \vvvl_1 \vvvd_1+ \vvvd_1 \vvvl_1}.
 \eal
We note that \eqref{curvi_lap_delta} gives $\Lapl \vvvd_2 \approx \ppsqr_\zs \vvvd_2$, and that using \eqref{cont_navier_second_short_a} the divergence term is rewritten as $-\beta\etaO \nablabf\big(\div\vvvd_2\big) = \beta\nuO\grad\avr{\vvvd_1\scap\grad\rho_1}= \frac{\beta \etaO \omega}{\cOsqr} \nablabf\avr{\vvvd_1 \scap \big(\ii\vvvl_1 \big)}$. This term is neglected as it is smaller by a factor $\Gamma$ compared to the other divergence term. In the BL18 model, the short-range 2nd-order pressure field $\pd_2$ was also neglected since the perpendicular velocity of the wall was assumed to be small. However, a major point in the BL25 model is that we relax this assumption and keep $\pd_2$. Thus, we end with two source terms for $\vvvd_2$, which therefore is split in two corresponding terms,
 \bsubal{Laplvd2Final}
 \eqlab{v2dSplit}
 \vvvd_2 &= \vvvddp + \vvvddv,
 \\ \nn
 &\;\text{where }\; \vvvddp(\xs,\ys,\infty) = \vvvddv(\xs,\ys,\infty) = \zerovec,
 \\
 \eqlab{Laplvddp}
 \ppsqr_\zs \vvvddp &= \frac{1}{\etaO}\grad \pd_2,
 \\
 \eqlab{Laplvddv}
 \ppsqr_\zs \vvvddv &= \frac{1}{\nuO} \div \avr{\vvvd_1 \vvvd_1+ \vvvl_1 \vvvd_1+ \vvvd_1 \vvvl_1}.
 \esubal

An analytical expression for $\vvvd_2$ is derived by separating the 1st-order fields in the source terms into the parallel coordinates $(\xi,\eta)$ and the perpendicular coordinate $\zs$,

 \bsubal{A1B1separation}
 \AAA_1 &= \ann(\zeta) \AAAs_1(\xi,\eta),
 \\
 \BBB_1 &= \bnn(\zeta)\BBBs_1(\xi,\eta),
 \\
 \timeav{\AAA_1}{\BBB_1} &= \timeav{a\AAAs_1}{b\BBBs_1} =
 \timeav{ab^*\AAAs_1}{\BBBs_1},
 \\ \nn
 & \quad \text{ with }\; \timeav{A_1}{B_1} = \avr{A_1B_1},
 \esubal
where in the last expression, we have introduced the comma-separated notation $\timeav{A_1}{B_1} = \avr{A_1B_1}$ to ensure clarity when moving pre-factors back and forth between $A_1^0$ and $B_1^0$, such as when collecting all $\zs$-dependence in the product $a(\zs)b^*(\zs)$. The appearance of the simplified Laplacian~\eqnoref{curvi_lap_delta}, $ \Lapl \gd \approx \ppperpsqr\gd$, in the boundary-layer equations~\eqnoref{equMotion2}, implies, that we need to integrate terms $\timeav{\AAA_1}{\BBB_1}$ with respect to $\zs$. To this end, as in the BL18 model, we introduce the recursive $n$-fold integrals $\intnab nab(\zs)$ for integer values $n=1,2,3, \ldots$, and for the extensions $n= 0$ (no integration) and $-n$ (the $n$th derivative),
 \bsubal{def_intnab} 
 \eqlab{abInfinity}
 \intnab 0ab (\zs) &= a(\zs)\: b(\zs)^*, \quad
 \lim_{\zs \rightarrow \infty} \big[a(\zs)\: b(\zs)^*\big] \rightarrow 0,
 \\
 \eqlab{intnabIntegral}
 \intnab {n}ab (\zs) &=  \int_\infty^\zs
 \dm\zs_n \intnab {n-1}ab (\zs_n), \quad  n=1,2,3, \ldots,
 \\
 \intnab {-n}ab (\zs) &= \partial_\zs^n \big[a(\zeta)\: b(\zeta)^*], \quad  n=1,2,3, \ldots,
 \\
 \intnab {n}ba (\zs) &= \big[\intnab {n}ab (\zs)\big]^*, \quad  \pm n =0,1,2,3, \ldots,
 \\
 \timeav{\AAA_1}{\BBB_1} &=  \timeav{\intnab {0}ab (\zs)\:\AAAs_1(\xi,\eta)}{\BBBs_1(\xi,\eta)}.
 \esubal

We also need the coordinate-separated expressions for the $i$th component $\big[\div\timeav{\AAA_1}{\BBB_1}\big]_i$ of the single-divergence $\div\timeav{\AAA_1}{\BBB_1}$,
 \bsub
 \bal \eqlab{single_div_ABi}  
 &  \big[\div \timeav{\AAA_1}{\BBB_1}\big]_i
= \div \timeav{\intnab {0}ab (\zs) \Ann_{1i}^0}{\BBBs_1}
 \\ \nn
 & = \pardiv \timeav{(\intnab {0}ab (\zs) \Ann_{1i}^0}{\BBB_{1\Vert}^{0}}
 + \timeav{\intnab {-1}ab (\zs) \Ann_{1i}^0}{\Bnn_{1\perp}^0}.
 \eal
with tangential and normal components
 \bal
 \eqlab{single_div_ABparallel}  
 \big[\div &  \timeav{\AAA_1}{\BBB_1}\big]_\parallel
 \\ \nn
 & = \pardiv \timeav{(\intnab {0}ab (\zs) \AAA_{1\parallel}^0}{\BBB_{1\Vert}^{0}}
 + \timeav{\intnab {-1}ab (\zs) \AAA_{1\parallel}^0}{\Bnn_{1\perp}^0},
 \\
 \eqlab{single_div_ABzeta}  
 \big[\div & \timeav{\AAA_1}{\BBB_1}\big]_\zs
 \\ \nn
 & = \pardiv \timeav{(\intnab {0}ab (\zs) \Ann_{1\zs}^0}{\BBB_{1\Vert}^{0}}
 + \timeav{\intnab {-1}ab (\zs) \Ann_{1\zs}^0}{\Bnn_{1\perp}^0},
 \eal
and for the double-divergence $\div\big[\div\timeav{\AAA_1}{\BBB_1}\big]$,
 \bal \eqlab{double_div_AB}  
 \div &  \big[\div \timeav{\AAA_1}{\BBB_1}\big]
 \\ \nn
 & = \pp_i\big[\pardiv \timeav{(\intnab {0}ab (\zs) \Ann_{1i}^0}{\BBB_{1\Vert}^{0}}
 + \timeav{\intnab {-1}ab (\zs) \Ann_{1i}^0}{\Bnn_{1\perp}^0}\big]
 \\\nn
 & = \pardiv\big[\pardiv\timeav{\intnab {0}ab (\zs)\AAA_{1\Vert}^0}{\BBB_{1\Vert}^0}\big]
  + \timeav{\intnab {-2}ab (\zs) \Ann_{1\zeta}^0}{\Bnn_{1\zeta}^0}
 \\ \nn
 & \quad
 +\pardiv \big[\timeav{\intnab {-1}ab (\zs) \AAA_{1\Vert}^0}{\Bnn_{1\zeta}^0}
  + \timeav{\intnab {-1}ab (\zs) \Ann_{1\zeta}^0}{\BBB_{1\Vert}^0}\big].
 \eal
 \esub

Using the solution~\eqnoref{vd_sol} for $\vvvd_1$ and a 1st-order Taylor-expansion in $\zs$ of $\vvvl_1$ valid in the boundary layer $\zs \ll d$, we obtain the explicit coordinate-separated form for the relevant velocity fields in the boundary layer,\\[-5mm]
 \bsubalat{vd1vl1BL}{2}{
 \eqlab{vd1BL} 
 \vvvd_1 & =q(\zeta)\:\vvvds_1(\xi,\eta), &\quad &
 \text{with }\; q(\zeta) = \ee^{\ii\ks\zs},
 \\
 \eqlab{vl1BL} 
 \vvvl_1 &= \vvvls_1 + \zs\:\pp_\zeta\vvv^{d0}_1, &\quad &
 \text{with }\; \pp_\zs\vvv^{d0}_1 = \lim_{\zeta\rightarrow 0}\pp_\zeta\vvv^d_1.
 }\esubalat
We now see that the functions $a(\zeta)$ and $b(\zeta)$ of \eqref{A1B1separation} in the source terms of \eqref{equMotion2} are either $q(\zeta)$, $\zeta$, or unity, and that each product $a(\zs)b(\zs)$ contains at least one decaying factor $q(\zs)$, so the zero-at-infinity boundary condition~\eqnoref{abInfinity} is fulfilled. On the surface $\zs = 0$, $\intnab nab (\zs)$ becomes $\intBCnab nab = \intnab nab (0)$, which leads to the explicit values,
 \balat{3}{
 \eqlab{IabnBC} 
 \intBCnab {-1}qq   &= 0, \quad &
 \intBCnab {-1}q1   &= \frac{-1+\ii}{\delta}, \quad &
 \intBCnab {-1}q\zs &= 1,
 \\
 \nn
 \intBCnab {0}qq    &= 1, \quad &
 \intBCnab {0}q1    &= 1, \quad &
 \intBCnab {0}q\zs  &= 0,
 \\
 \nn
 \intBCnab {1}qq    &= -\frac12 \delta, \quad &
 \intBCnab {1}q1    &= -\frac{1+\ii}{2} \delta, \quad &
 \intBCnab {1}q\zs  &= -\frac{\ii}{2} \delta^2,
 \\
 \nn
 \intBCnab {2}qq    &= \frac14 \delta^2, \quad &
 \intBCnab {2}q1    &= \frac{\ii}{2} \delta^2, \quad &
 \intBCnab {2}q\zs  &= -\frac{1-\ii}{2} \delta^3,
 \\
 \nn
 \intBCnab {3}qq    &= -\frac18 \delta^3, \quad &
 \intBCnab {3}q1    &= \frac{1-\ii}{4} \delta^3,  \quad &
 \intBCnab {3}q\zs  &= \frac{3}{4} \delta^4,
 \\
 \nn
 \intBCnab {4}qq    &= \frac1{16} \delta^4, \quad &
 \intBCnab {4}q1    &= -\frac14 \delta^4,   \quad &
 \intBCnab {4}q\zs  &= -\frac{1+\ii}{2} \delta^5.
 }\ealat

We can now find the analytical solution for $\vvvd_2$ to be used in the slip boundary condition~\eqnoref{decomp_bc} for the bulk streaming velocity $\vvvls_2$ at the wall $\zeta = 0$. As shown in \appref{BC2ndOrderAppendix},  $\vvvd_2$ is found by integrating \eqref{Laplvd2Final} twice from $\zs = \infty$, where it is zero, to $\zs$ using that: (1) the right-hand-side of \eqref{Laplvd2Final}, called $f^\mr{rhs}$, can be separated in the in-plane and perpendicular coordinates $(\xs,\ys)$ and $\zs$, respectively, (2) $f^\mr{rhs}$ forms a linear superposition of functions $f^\mr{rhs}_\alpha(\xs,\ys) \intnab{n_\alpha}ab(\zs)$, $\alpha = 1, 2, \ldots, M$, and (3) the functions $\intnab{n_\alpha}ab(\zs)$ have the special properties summarized in \eqref{def_intnab}, notably the integration property~\eqnoref{intnabIntegral} that changes a $\zs$-integration into a simple increase by unity of the index $n_\alpha$. The structure of the computation of $\vvvd_2$ is simply (see details in \appref{BC2ndOrderAppendix}),
 \bsubal{v2deltaForm}
 \ppsqr_\zs \vvvd_2 &= \sum_{\alpha=1}^M f^\mr{rhs}_\alpha(\xs,\ys)\: \intnab{n_\alpha}ab(\zs),\;
 \text{ which implies}
 \\
 \vvvd_2 &= \sum_{\alpha=1}^M f^\mr{rhs}_\alpha(\xs,\ys)\: \intnab{n_\alpha+2}ab(\zs).
 \esubal

The surface value $\vvvds_2$ of the  boundary-layer field $\vvvd_2$ is derived in \appref{BC2ndOrderAppendix} in terms of its tangential and normal components, including the pre-factor$\frac{1}{\nuO} = \frac{2}{\omega \delta^2}$ and by substituting the bulk factors $\intnab nab(\zeta)$ with their surface values $\intBCnab nab$ given in \eqref{IabnBC}, with $\frac{2}{\omega \delta^2}\intBCnab nqq, \frac{2}{\omega \delta^2}\intBCnab n1q  \propto \delta^{n-2}$ and $\frac{2}{\omega \delta^2}\intBCnab n{\zeta}q  \propto \delta^{n-1}$. The result is
 \bsuba{vddFinal}
 \bal
 \eqlab{vddFinalSum0}
 \vvvds_2 &=  \vvvddspar + \vddsperp\:\een_\zs,
 \\
 \eqlab{vddFinalPar}
 \vvvddspar &= \frac{2}{\omega \delta^2} \Big\{
 -\pargrad \Big[\timeav{\intBCnab 2qq \vdsperp}{\vdsperp}
 \\ \nn
 &
 \qquad\qquad\qquad
 +\timeav{\intBCnab 21q \vlsperp}{\vdsperp}
 +\timeav{\intBCnab 2q1 \vdsperp}{\vlsperp}\Big]
 \\ \nn
 &
 +\pardiv \timeav{\intBCnab 2qq \vvvdspar}{\vvvdspar} 
 +\timeav{\intBCnab 1qq  \vvvdspar}{\vdsperp}  
 \\ \nn
 &+
 \pardiv \timeav{\intBCnab 21q \vvvlspar}{\vvvdspar}  
 +\timeav{\intBCnab 11q  \vvvlspar}{\vdsperp}  
 \\ \nn
 &+
 \pardiv \timeav{\intBCnab 2q1  \vvvdspar}{\vvvlspar} 
 +\timeav{\intBCnab 1q1 \vvvdspar}{\vlsperp} 
 \\ \nn
 &
 +\timeav{\intBCnab 1{\zs}q \ppperp  \vvvlspar}{\vdsperp} 
 +\timeav{\intBCnab 1q{\zs} \vvvdspar}{\ppperp\vlsperp}
 \Big\},
 \\
 \eqlab{vddFinalPerp}
 \vddperpO &= -\frac{2}{\omega \delta^2} \Big\{
 \\ \nn
 &
 +\pardiv\big[\pardiv \timeav{\intBCnab 3qq \vvvdspar}{\vvvdspar} 
 +\timeav{\intBCnab 2qq \vvvdspar}{\vdsperp}\big] 
 \\ \nn
 &
 +\pardiv \big[\pardiv \timeav{\intBCnab 31q \vvvlspar}{\vvvdspar}   %
 +\timeav{\intBCnab 21q \vvvlspar}{\vdsperp}\big] 
 \\ \nn
 &
 +\pardiv \big[\pardiv \timeav{\intBCnab 3q1 \vvvdspar}{\vvvlspar} 
 + \timeav{\intBCnab 2q1 \vvvdspar}{\vlsperp} \big]  
 \\ \nn\
 &
 +\pardiv \big[\timeav{\intBCnab 2{\zs}q \ppperp\vvvlspar}{\vdsperp}
 +\timeav{\intBCnab 2q{\zs} \vvvdspar}{\ppperp \vlsperp} \big]  
 \Big\}.
 \eal
 \esuba
Here, terms to the two lowest powers in the small length scale $\delta$ are kept, which for \eqref{vddFinalPar} is $\delta^{-1}$ and $\delta^0$, and for \eqref{vddFinalPerp} is $\delta^0$ and $\delta^1$. These expressions for $\vvvds_2$ are central in the BL25 model. In BL18 only the lowest power $\delta^{-1}$ in \eqref{vddFinalPar} and $\delta^0$ in \eqref{vddFinalPerp} was kept.

\subsection{Long-range bulk streaming}
\seclab{LongRangeStreaming}
The governing equations for the 2nd-order bulk streaming follow from the long-range part of
\eqref{cont_navier_second}, and they are the same as in the BL18 model,
 \bsubal{cont_navier_second_long} 
 \eqlab{cont_navier_second_long_a} 
 \rho_0\div\vvvl_2	&=\div\avr{\rho_1\vvvl_1},
 \\ \nn
 \eqlab{cont_navier_second_long_b} 
 \div\sigmabfl_2 &= \rho_0\div\!\avr{\vvvl_1\vvvl_1}
 \\
 &= \rho_0\avr{\vvvl_1\cdot\grad\vvvl_1} + \rho_0\avr{\vvvl_1\:\div \vvvl_1},
 \\
 \eqlab{divsigmal2} 
 \text{with }\div \sigmabfl_2 &= -\nablabf\big(\pL_2-\beta\eta_0\div\vvvl_2\big)
 + \etaO \Lapl \vvvl_2.
 \esubal
As shown in Ref.~\cite{Bach2018}, the source terms in these equations can be evaluated using \eqsref{Helmholtz_decomp}{First-order_final},
 \bsubal{cont_navier_second_long_simple} 
 \eqlab{cont_navier_second_long_simple_a} 
 \div\vvvl_2 &= \Gamma\: \frac{\kO}{2\cO}\:\big| \vvvl_1 \big|^2 \approx 0,
 \\
 \eqlab{cont_navier_second_long_simple_b} 
 \etaO\Lapl\vvvl_2 &= \grad \ptL_2 -
 \frac{\Gamma\omega}{\cO^2}\avr{\pL_1\vvvl_1},
 \\
 \ptL_2 &= \pL_2 -\big(\tfrac14 \kapO\: |\pL_1|^2 - \tfrac14\rho_0\: |\vvvl_1|^2\big),
 \esubal
where we have introduced the excess pressure $\ptL_2$, which is the 2nd-order pressure $\pL_2$ minus the time-averaged acoustic Lagrangian density.

Next, we express the 2nd-order boundary condition~\eqnoref{decomp_bc}, $\vvvls_2 = -\vvvds_2 - \avr{(\sss_1\cdot\grad)\vvv_1}|_{\zs=0}$ in terms of the 1st-order surface values $\vvvwall_1$ and $\vvvl_1$ of the wall velocity and the bulk acoustic velocity, respectively, noting that by \eqref{vvvds_sol} we have $\vvvds_1 = \vvvwall_1 - \vvvls_1$. This has already been achieved for the surface value $\vvvds_2$ of the boundary streaming in \eqref{vddFinal}, only the Stokes-drift term $\vvv^\mr{sd}_2 = \avr{(\sss_1\cdot\grad)\vvv_1}|^\notop_{\zs=0}$ needs to be treated here. We use \eqref{bc} $\sss_1 = \frac{\ii}{\omega}\vvvwall_1$, and \eqsref{AAApar}{vd1vl1BL} to compute in-plane and normal derivatives, and obtain
 \bal
 \eqlab{StokesDriftWall}
 \vvv^\mr{sd}_2 &= \avr{(\ssn_1\!\cdot\!\grad)\vvn_1}|_{\zs=0}
 \\ \nn
 & = \frac{1}{\omega}\Big[\timeav{\ii(\vvvwall_1\cdot\grad)}{\vvvl_1 + \vvvd_1}\Big]^\notop_{\zs=0}
 \\ \nn
 &= \frac{1}{\omega}\Big[
   \timeav{\ii\vvvwall_1\cdot\:}{\grad \vvvls_1+\pargrad \vvvds_1}
 + \timeav{\vwallperp}{\ks\vvvds_1} \Big].
 \eal
The final form of the slip boundary condition \eqnoref{decomp_bc} for $\vvvls_2$, is obtained by combining \eqsref{vddFinal}{StokesDriftWall},
 \beq{v20Final}
 \vvvls_2 = -(\vvvds_2 + \vvv^\mr{sd}_2)
 = -(\vvvddspar + \vvv^\mr{sd}_{2\parallel}) -  (\vddsperp+v^\mr{sd}_{2\zeta}) \:\een_\zs.
 \eeq
In the BL25 model, the 2nd-order steady bulk streaming $\vvvl_2$ appears as a weakly compressible Stokes flow governed by~\eqref{cont_navier_second_long_simple}, which is driven by the weak body force $-\frac{\Gamma\omega}{\cO^2}\avr{\pL_1\vvvl_1}$ and by the slip-boundary condition~\eqnoref{v20Final}. All the driving terms have the form of time-averaged products of the 1st-order fluid pressure and velocity fields $\pI$ and $\vvvI$ and the 1st-order solid displacement field $\uuu_1$ with wall velocity $\vvvwall_1$ as described in \secsref{pert_theory}{fl_first_order}, respectively. In the following section we describe how this model is implemented in Comsol Multiphysics for numerical solution.

%
%

\section{Acoustofluidic device modeling}
\seclab{device_model}
We introduce two polymer-based model systems, a 3D device driven by a thin-film transducer and a 2D device driven by a bulk transducer, as sketched in \figref{device_sketch} with the specific geometrical parameters listed in \tabref{device_dimensions}.  The top electrode of the transducer in both models are split by a groove to ensure an efficient excitation of a selected antisymmetric resonance mode.

\begin{table}[t]
\centering
{
\caption{\tablab{device_dimensions} The geometrical parameters for the 3D and 2D polymer-based devices considered here, see also \figref{device_sketch}.}
\begin{ruledtabular}
\begin{tabular}{crccr}
 Symbol            & Value       &   \hspace*{8.0em}     &  Symbol          & Value \\
 \hline
   \multicolumn{5}{l}{\textit{The 3D device driven by a} \AlScN\ \textit{thin-film transducer\upspace}}\\
 $L_\sl$ & $500~\SImum$      &  & $L_\fl$ & $300~\SImum$ \\
 $W_\sl$ & $475~\SImum$      &  & $W_\fl$  & $275~\SImum$ \\
 $H_\sl$ & $320~\SImum$      &  & $H_\fl$  & $120~\SImum$ \\
 $H_\pz$ & $2~\SImum$        &  & $T_\wl$ & $100~\SImum$ \\ \hline
  \multicolumn{5}{l}{\textit{The 2D device driven by a bulk PZT transducer \upspace}}\\
 $W_\sl$   & $2750~\SImum$ &  & $W_\fl  $ & $392~\SImum$ \\
 $H_\sl$   & $650~\SImum$  &  & $H_\fl  $ & $351~\SImum$ \\
 $W_\pz$     & $3500~\SImum$ &  & $W_\gr$ & $150~\SImum$ \\
 $H_\pz$     & $500~\SImum$ &  & $H_\gr$ & $100~\SImum$ \\
 $H_\mr{el}$ & $9~\SImum$    &  & $H_\mr{gl}$ & $20~\SImum$
\end{tabular}
\end{ruledtabular}
}
\end{table}

\begin{figure}[t]
\centering
\includegraphics[width=\columnwidth]{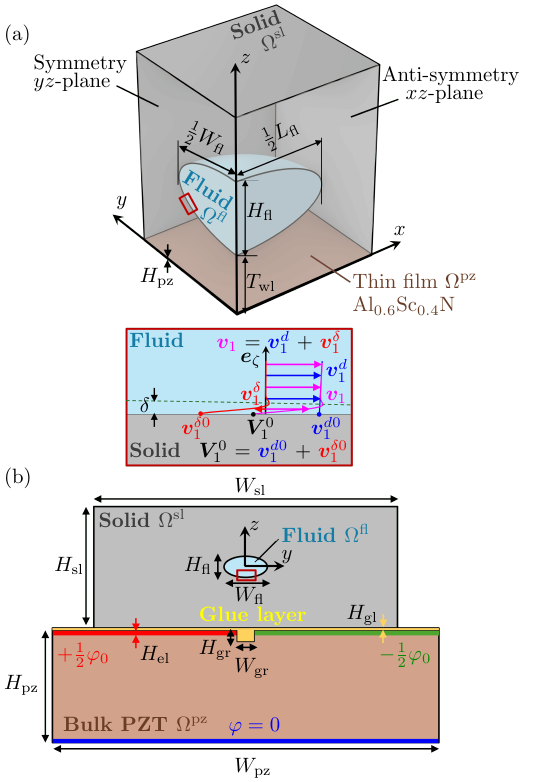}
\caption[]{\figlab{device_sketch}
Two acoustofluidic device models used for numerical validation of the BL25 model against the Full model, see parameter values in \tabref{device_dimensions}. (a) The small sub-mm 3D model consisting of an ellipsoidal fluid domain embedded in a solid with a thin-film \AlScN\ transducer attached on its lower surface. (b) The 2D mm-sized model consisting of an elliptic fluid domain embedded in a solid, which is attached to a bulk PZT transducer via a glue layer. The inset is a zoom-in of the fluid-solid interface that illustrates the Helmholtz decomposition~\eqnoref{Helmholtz_decomp} $\vvv_1 = \vvvl_1+\vvvd_1$ of the acoustic velocity field.}
\end{figure}

\subsection{The 3D model with a thin-film transducer}
\seclab{device3D}

The 3D model is made as small as possible to allow for simulations using the Full model within the available 600-GB-RAM memory capacity of our computer system, while still containing all essential elements of an acoustofluidic device: a fluid cavity (fl) embedded in a solid (sl), which is driven by an attached piezoelectric transducer (pz). Spurious corner effects are avoided by choosing an ellipsoidal fluid domain with principal axes $L_\fl \gtrsim W_\fl \gtrsim H_\fl$, all less than $300~\SImum$. The dimensions of the box-shaped solid of length $L_\sl$, width $W_\sl$, and height $H_\sl$ are less than $500~\SImum$, which results in resonance frequencies in the relevant 1-2 MHz range. The elastic solid is the polymer poly(methyl methacrylate) (PMMA), and the smallest possible transducer is a thin-film \AlScN\ transducer of height $H_\pz \simeq 2~\SImum$, which covers the bottom surface of the solid as studied in Refs.~\cite{Steckel2021, Steckel2021b}.

Finally, by assuming the $yz$- and $xz$-plane to be a symmetry and an anti-symmetry plane, respectively, the computational domain is reduced to a quarter of the full domain. The boundary conditions presented in the previous sections therefore needs to be supplemented by the following (anti-)symmetry-plane conditions valid for each perturbation order,
 \bsuba{BC_sym_antisym}
 \balat{4}{
 \eqlab{BC_sym}
 \text{Symm}&\text{etry at $x=0$:}\hspace*{-30mm}  && && &&
 \\  \nn
 \pp_{x} \pI &= 0,\quad &
 \pp_x \vphn_1 &= 0,\quad &
 u_{1,x} &= 0,\quad &	
 \sigmasl_{1yx} & = \sigmasl_{1zx} = 0,
 \\ \nn
 \pp_{x} \pII &= 0,\; &
 &&
 \vl_{2x} &= 0,\quad &	
 \sigma^{d,\fl}_{2yx} & = \sigma^{d,\fl}_{2zx} = 0,
 \\[2mm]
 \eqlab{BC_sym_y0}
 \text{Symm}&\text{etry at $y=0$:}\hspace*{-20mm}  && && &&
 \\ \nn
 \pp_y\pII &= 0,\quad  &
 & \qquad\quad&
 \vl_{2y}& =0, \quad &	
 \sigma^{d,\fl}_{2xy} & = \sigma^{d,\fl}_{2zy} = 0,
 }\ealat
 \balat{4}{
 \eqlab{BC_antisym} 
 \text{Anti-s}&\text{ymmetry at $y=0$:}\hspace*{-20mm}  && && &&
 \\ \nn
 \pI &= 0,\;  &
 \vphn_1 &= 0, \; &
 \sigmasl_{1yy} &= 0,\;  &
 u_{1x} &= u_{1z} = 0,
 }\ealat
 \esuba

Finally, as there in the symmetry-reduced model are only a top and a bottom electrode on the transducer, the potential boundary condition \eqnoref{BCpzElec} becomes,
 \bsubalat{BCelec3D}{2}
 \vphn_1 &= +\tfrac12 \vphn_0,\quad &&\text{for } \rrr \in \pp\Omegapz_\mr{top},
 \\
 \vphn_1 &= 0,\quad && \text{for } \rrr \in \pp\Omegapz_\mr{bottom},
 \esubalat
The 20-nm-thick metal electrodes are not explicitly included in the model, but they appear indirectly through the potential values~\eqnoref{BCelec3D} on the transducer surfaces.

\subsection{The 2D model with a bulk transducer}
The 2D model is nearly the same as the cross section of the actual mm-sized bulk-transducer-driven polymer-based device studied in Ref.~\cite{Lickert2021}. However, to avoid spurious corner effects in the numerical validation, we have changed the original rectangular fluid domain to a smooth elliptical domain with major axis $W_\fl$ and minor axis $H_\fl$. The rectangular solid and bulk lead-zirconate-titanate (PZT) transducer has width $\times$ height given by $W_\sl \times H_\sl$ and $W_\pz \times H_\pz$, respectively. The glue layer between the transducer and the solid has the thickness $H_\mr{gl}$, the thickness of the electrodes is $H_\mr{el}$, and the glue-filled grove in the top electrode has width $W_\gr$ and height $H_\gr$. Note that in this model there are three isotropic elastic solid domains with corresponding Cauchy momentum equations~\eqnoref{CauchyEqu}: the PMMA wall (sl = pm), the glue layer (sl = gl), and the silver electrodes (sl = el), and that the boundary conditions~\eqnoref{BCslpz} according to \figref{device_sketch} instead of only $\pp\Omega_\sl^\pz$ now include three types of interfaces $\pp\Omega_\mr{el}^\pz$, $\pp\Omega_\mr{gl}^\pz$, and $\pp\Omega_\mr{gl}^\mr{pm}$.

Finally, the potential boundary condition \eqnoref{BCpzElec} is specified by the anti-symmetric form around the $y=0$ line of the voltages applied to each of the three electrodes of the 2D device, the top-left, top-right, and bottom electrode,\negspace
 \bsubalat{BCelec2D}{2}
 \vphn_1 &= +\tfrac12 \vphn_0,\quad &&\text{for } \rrr \in \pp\Omegapz_\mr{top-left},
 \\
 \vphn_1 &= -\tfrac12 \vphn_0,\quad &&\text{for } \rrr \in \pp\Omegapz_\mr{top-right},
 \\
 \vphn_1 &= 0,\quad && \text{for } \rrr \in \pp\Omegapz_\mr{bottom},
 \esubalat

%
%

\section{Numerical implementation in COMSOL Multiphysics}
\seclab{COMSOL_impl}
Following the previous works from our group, the governing partial differential equations (PDE) are implemented in the finite-element method software COMSOL Multiphysics \cite{Comsol62} using the \qmtt{Weak Form PDE Interface} \cite{Bach2018, Skov2019} (see a sample script in the Supplemental Material of Ref.~\cite{Joergensen2023}). All simulations consist of two study steps: In step 1, the 1st-order acoustic equations are solved, and the result is passed on to step 2, in which the time-averaged 2nd-order equations are solved using the 1st-order solutions in the different source terms.

In the Full model, the acoustic fields are the pressure $\pI$ and the full acoustic velocity $\vvvI$ in the fluid, the displacement $\uun_1$ in the solids, and the electric potential $\vphn_1$ in the transducer. The governing equations of these fields are \eqssref{CauchyEqu}{transducerEq}{cont_navier_1}, and the corresponding boundary conditions are \eqssref{BCsolid}{BCpiezo}{BCelec2D}. The time-averaged fields are only the pressure $\pII$ and the full streaming field $\vvvII$ in the fluid, with the governing Stokes-flow equation and Stokes-drift boundary condition both given in \eqref{cont_navier_second}.

In the BL25 model, the acoustic field is just the pressure $\pI$ in the fluid, the displacement $\uun_1$ in the solids, and the electric potential $\vphn_1$ in the transducer. The governing equations of these fields are \eqssref{CauchyEqu}{transducerEq}{Helmholtz_p1}, and the corresponding boundary conditions are \eqsref{BCsolid}{BCpiezo} as well as the self-consistent, effective, boundary-layer conditions for the pressure and the  stress, \eqsref{p1_bc}{stress1_bc}. The time-averaged fields are only the excess pressure $\ptL_2$ and the long-range bulk streaming field $\vvvl_2$ in the fluid, with the governing weakly-compressional-Stokes-flow equation~\eqnoref{cont_navier_second_long_simple} and the effective boundary-layer velocity slip condition given by \eqref{v20Final} in combination with \eqsref{vddFinal}{StokesDriftWall}. An important technical note regards third derivatives in COMSOL, which appear several places in the boundary condition~\eqnoref{vddFinal}, such as $\pp_z\pp_y v^d_{1x} \propto \pp_z\pp_y \pp_x p_1$. In COMSOL, such a term is set to zero, as three consecutive derivatives of any field are not allowed. Consequently,  we introduce an auxiliary vector field $\GGn_1$ equal to the pressure gradient $\grad\pI$ such that its $k$th component is $G_{1k} = \pp_k\pI$. Following \eqref{v1dgradp1}, we set $\vvvl_1 = -[\ii(1-\ii\Gamma)/\omega\rhoO]\GGn_1$, which in COMSOL has nonzero second derivatives, such as $\pp_z\pp_y v^d_{1x} \neq 0$.

To understand the following details in the COMSOL implementation, it is relevant to mention that in COMSOL, the partial $x$-derivative $\pp_x f$ of a scalar field $f$ is written by appending a lower-case \qmtt{x} to \qmtt{f} as  \qmtt{fx}. Thus the Cartesian coordinates of the gradient $\grad f$ is written as  \qmtt{(fx,fy,fz)}. To avoid ambiguities, the Cartesian vector components of a vector $\vvv$ are represented by upper-case letters such as \qmtt{(vX,vY,vZ)}. The partial $x$-derivative $\pp_x \vnn_x$ of the component $\vnn_x$ is thus written  \qmtt{vXx}.

In the COMSOL implementation of the effective boundary conditions of the BL25 model, the local tangential and normal unit vectors $\een_\xs = \ttn_1$, $\een_\ys=\ttn_2$, and $\een_\zs = \nnn$, the tangential derivatives $\pargrad$, and the time-average $\timeav{a_1}{b_1}$ of 1st-order products $a^\notop_1 b^\notop_1$ play a crucial role. In COMSOL, the local tangential and normal vectors have the Cartesian components \qmtt{(t1X,t1Y,t1Z)}, \qmtt{(t2X,t2Y,t2Z)}, and \qmtt{(nX,nY,nZ)}, respectively. As we work only in the limit of weakly-curved interfaces as described in \secref{WeaklyCurvedInterfaces}, all $\zeta$-derivatives of normal and tangential vectors are neglected. For example, the $\zeta$-derivative $\pp_\zs \vvn_\zs$ of the  of the $\zeta$-component of the interface vector $\vvv^0$ is computed as $\pp_\zs (\nnn\cdot\vvv^0) \approx \nnn\cdot\pp_\zs \vvv^0 = n^{{}}_i(n^{{}}_k\pp_k) v^{0_{}}_i$, which in COMSOL is the nine-term sum \qmtt{nX*nX*vXx + nX*nY*vXy + \ldots + nZ*nZ*vZz}. The tangential derivatives are computed without approximations using the COMSOL operator \qmtt{dtang()}, and the time averages are computed using the COMSOL operator \qmtt{realdot()}. To avoid the spurious imaginary parts that sometimes appear, we explicitly take the real part, \qmtt{real(realdot())} of all \qmtt{realdot()} terms.

Since we cannot use $\delta$ as a superscript in COMSOL scripts, we introduce the superscript changes: $\delta \rightarrow $ \qmtt{d} (for delta), $d \rightarrow $ \qmtt{f} (for fluid bulk), and $\mr{sl} \rightarrow $ \qmtt{s} (for solid). In COMSOL at the fluid-solid interface $\zs = 0$, the Cartesian coordinates of the wall velocity $\vvvwall_1$, the bulk fluid velocity $\vvvl_1$, and the boundary-layer velocity $\vvvd_1$ are written as \qmtt{(vs1X,vs1Y,vs1Z)}, \qmtt{(vf1X,vf1Y,vf1Z)}, and \qmtt{(vd1X,vd1Y,vd1Z)}, where the latter according to $\vvvds_1 = \vvvwall_1-\vvvls_1$ is computed as  \qmtt{(vs1X-vf1X, vs1Y-vf1Y, vs1Z-vf1Z)}. As mentioned above,
$\GGn_1$ is introduced as an auxiliary field equal to $\grad p_1$ by the weak form definition  \qmtt{test(GK)*(GK-p1k)} for K = X, Y, Z and k = x, y, z. The bulk velocity and its gradient are correspondingly defined as \qmtt{vf1K = (1-i*Gam0)/(i*omega*rho0)*GK} and \qmtt{vf1KN = (1-i*Gam0)/(i*omega*rho0)*GKn} with K, N = X, Y, Z and n = x, y, z. All tangential derivatives after coordinate $k$ of component $K$ of the wall velocity $\vvvwall_1$ is computed as \qmtt{vs1Ktk = dtang(vs1K,k)} with $\mathtt{K = X, Y, Z}$ and $\mathtt{k = x, y, z}$, and similar for $\vvvls_1$ and $\vvvds_1$. Here, \qmtt{Ktk} in the name \qmtt{vs1Ktk} is read as \qm{the $K$th component's tangential $k$-derivative}.

Using this notation, we implement all the governing equations and boundary conditions of the BL25 model. In the following we show, how the 1st-order boundary conditions are implemented, and then refer to \appref{COMSOLimplem} for a more detailed description including the elaborate 2nd-order boundary conditions. The 1st-order pressure boundary condition~\eqnoref{p1_bc} is written as
\qmtt{i*omega*rho0/(1-i*Gam0)*(nX*vs1X+nY*vs1Y + nZ*vs1Z - ii/ks*divparvs10)
 - ii/ks * (kc\^{}2*p1 + p1zetazeta)}, where \qmtt{divparvs10} is
the tangential divergence $\pardiv \vvvwallpar$ written as \qmtt{vs1XtX+vs1YtY+vs1ZtZ} and \qmtt{p1zetazeta} is the
the 2nd-order $\zeta$-derivative $\ppsqr_\zeta p_1 =(\nnn\cdot\grad)(\nnn\cdot\grad) p_1$ written as \qmtt{= nX*nX*p1xx
+ 2*nX*nY*p1xy + 2*nX*nZ*p1xz + nY*nY*p1yy + 2*nY*nZ*p1yz  + nZ*nZ*p1zz}. Similarly, the $k$-component $\een_k\cdot \sigmabf^\sl_1\cdot\een_\zs$ of the 1st-order stress boundary condition \eqnoref{stress1_bc} is written as
\qmtt{-p1*nK + i*ks*eta0 *(vd1K+nK*vd1zeta)}, where $\mathtt{K = X, Y, Z}$ and \qmtt{vd1zeta = vd1X*nX+vd1Y*nY+vd1Z*nZ}.
%

%
%

\section{Numerical Validation}
\seclab{numValidation}
To validate the BL25 model numerically, we have simulated the 3D and 2D polymer-based acoustofluidic devices with elliptical fluid channels shown in \figref{device_sketch} and with the material parameters tabulated in \tabref{material_values}. As described below, the BL25-model results were successfully first tested for internal consistency by performing a mesh-convergence test, and then they were compared with results obtained by direct numerical simulations of the same devices using the Full model. In 3D, we find that using the Full-model for direct numerical simulation of acoustofluidic devices of typical experiments \citep{Antfolk2014, Hoque2024} is out of the question, as it would required more than 1,000~GB of memory making it harder to simulate even using the DTU High Performance Computing (HPC) cluster at our disposal. Hence, we consider the tiny, yet physically realistic, device described above, which, although not of practical use, at least allows us to compare the BL25 model solutions with the Full model solutions. We remark that even with this tiny device, we are close to the limit of the finest mesh resolution that we can obtain when running the Full model on the DTU HPC cluster. In the 2D case, we are not near this mesh resolution limit, and we compared both the BL25 and the BL18 model to the Full model.

Qualitative validation in the form of plots, are for selected field components $g$ supplemented by the quantitative relative deviation $\eps^\notop_2(g^\mr{B},g^\mr{F})$ of the BL-model field $g^\mr{B}$ from the Full model field $g^\mr{F}$ using the $L^2$-norm on the  domain~$\Omega$~\cite{Muller2012},
 \beq{L2norm}
 \eps^\notop_2(g^\mr{B},g^\mr{F})  =
 \frac{\norm{g^\mr{B}-g^\mr{F}}_2}{\norm{g^\mr{F}}_2}, \text{ with }
 \norm{g}_2 = \sqrt{\int_\Omega|g|^2\:\dm V}.
 \eeq

\begin{table}[t!]
\centering
{
\caption{\tablab{material_values} Parameters values at $25~\SICel$ used in the numerical simulation of fluids (water, iodixanol-water, glycerol-water), solids (PMMA, NOA86H glue), and transducers (PZT Pz27 and \AlScN). For the isotropic solids  $C_{12} = C_{11} - 2C_{44}$, and for the transducers $C_{12}  = C_{11} - 2C_{66}$.}
\begin{ruledtabular}
\begin{tabular}{lcrl}
 Parameter &  Symbol  & Value & Unit
 \\
 \hline
 \multicolumn{4}{l}{\textit{Water} \citep{Muller2014}}\upspace \\
 Mass density & $\rhofl$ & $997.05$ & $\SIkgpcm$ \\
 Speed of sound & $\cfl$ & $1496.7$ & $\SImps$ \\
 Compressibility & $\kapfl$ & $447.7$ & $\SIpTPa$ \\
 Dynamic viscosity & $\etafl$ & $0.89$ & $\SImPas$ \\
 Bulk viscosity & $\etaflb$ & $2.485$ & $\SImPas$
 \\
 \hline
 \multicolumn{4}{l}{\textit{Iodixanol-water 16~\% V/V solution} \citep{Lickert2021, Muller2014} }\upspace\\
 Mass density & $\rhofl$ & $1050.0$ & $\SIkgpcm$ \\
 Speed of sound & $\cfl$ & $1482.3$ & $\SImps$ \\
 Compressibility & $\kapfl$ & $433.4$ & $\SIpTPa$ \\
 Dynamic viscosity & $\etafl$ & $1.474$ & $\SImPas$ \\
 Bulk viscosity & $\etaflb$ & $1.966$ & $\SImPas$
 \\
 \hline
  \multicolumn{4}{l}{\textit{Glycerol-water 99~\% V/V solution} \citep{Lickert2021, Cheng2008} }\upspace\\
 Mass density & $\rhofl$ & $1260.4$ & $\SIkgpcm$ \\
 Speed of sound & $\cfl$ & $1922.8$ & $\SImps$ \\
 Compressibility & $\kapfl$ & $214.6$ & $\SIpTPa$ \\
 Dynamic viscosity & $\etafl$ & $1.137$ & $\SIPas$ \\
 Bulk viscosity & $\etaflb$ & $0.790$ & $\SIPas$
 \\
 \hline
 \multicolumn{4}{l}{\textit{PMMA} \citep{Bode2022}}\upspace  \\
 Mass density & $\rhosl$ & $1162$ & $\SIkgpcm$ \\
 Elastic modulus & $C_{11}$ & $7.18 - \ii 0.183$ & $\SIGPa$ \\
 Elastic modulus & $C_{44}$ & $1.55 - \ii 0.111$ & $\SIGPa$
 \\
 \hline
 \multicolumn{4}{l}{\textit{Glue layer (NOA 86H)} \citep{Bode2022} }\upspace  \\
 Mass density & $\rhosl$ & $1300$ & $\SIkgpcm$ \\
 Elastic modulus & $C_{11}$ & $4.65 - \ii 0.51$ & $\SIGPa$ \\
 Elastic modulus & $C_{44}$ & $1.21 - \ii 0.12$ & $\SIGPa$
 \\
 \hline
 \multicolumn{4}{l}{\textit{PZT Pz27 bulk transducer} \citep{Bode2020}}\upspace  \\
 Mass density & $\rhopz$ & $7700$ & $\SIkgpcm$ \\
 Elastic modulus & $C_{11}$ & $124 - \ii 0.601$ & $\SIGPa$ \\
 Elastic modulus & $C_{12}$ & $ 77 - \ii 0.442$ & $\SIGPa$ \\
 Elastic modulus & $C_{13}$ & $ 80 - \ii 0.091$ & $\SIGPa$ \\
 Elastic modulus & $C_{33}$ & $119 - \ii 0.538$ & $\SIGPa$ \\
 Elastic modulus & $C_{44}$ & $20.4 - \ii 0.486$ & $\SIGPa$ \\
 Coupling constant & $e_{15}$ & $11.0$ & $\SIC\: \SIm^{-2}$ \\
 Coupling constant & $e_{31}$ & $-5.2$ & $\SIC\: \SIm^{-2}$ \\
 Coupling constant & $e_{33}$ & $16.3$ & $\SIC\: \SIm^{-2}$ \\
 Electric permittivity & $\ve_{11}$ & $982\;\epsO$ & $\SIF\: \SIm^{-1}$ \\
 Electric permittivity & $\ve_{33}$ & $804(1 - \ii 0.003)\;\epsO$ & $\SIF\: \SIm^{-1}$
 \\
  \hline
 \multicolumn{4}{l}{\AlScN\ \textit{thin-film transducer} \citep{Caro2015, Olsson2020, Steckel2021b}}\upspace  \\
 Mass density & $\rhopz$ & $3300$ & $\SIkgpcm$ \\
 Elastic modulus & $C_{11}$ & $313.8 - \ii 0.157$ & $\SIGPa$ \\
 Elastic modulus & $C_{12}$ & $150 - \ii 0.075$ & $\SIGPa$ \\
 Elastic modulus & $C_{13}$ & $139.2 - \ii 0.069$ & $\SIGPa$ \\
 Elastic modulus & $C_{33}$ & $197.1 - \ii 0.099$ & $\SIGPa$ \\
 Elastic modulus & $C_{44}$ & $108.6 - \ii 0.054$ & $\SIGPa$ \\
 Coupling constant & $e_{15}$ & $-0.317$ & $\SIC\: \SIm^{-2}$ \\
 Coupling constant & $e_{31}$ & $-2.653$ & $\SIC\: \SIm^{-2}$ \\
 Coupling constant & $e_{33}$ & $2.734$ & $\SIC\: \SIm^{-2}$ \\
 Electric permittivity & $\ve_{11}$ & $22.5(1 + \ii 0.0005)\;\epsO$ & $\SIF\: \SIm^{-1}$ \\
 Electric permittivity & $\ve_{33}$ & $22.5(1 + \ii 0.0005)\;\epsO$ & $\SIF\: \SIm^{-1}$
\end{tabular}
\end{ruledtabular}
\mbox{}\\[-10mm]}
\end{table}

A short summary of the following validation results is, that the BL25 model can predict the Full model results accurately for 1st- and 2nd-order fields in 3D and 2D, on and off resonance, and for hard glass-silicon and soft polymer devices. In contrast, the BL18 model (only tested in 2D) proved accurate only for 1st- and 2nd-order fields in hard glass-silicon devices run on resonance.

\subsection{Mesh convergence and computational times in 2D and 3D of the BL25 and Full model}
\seclab{MeshConv2D3D}

We have performed a mesh-convergence study for both the 3D and 2D model to determine the mesh, which gives a sufficient resolution for finding correctly converged solutions, see the details in \appref{MeshConvergence}. The mesh size is quantified by a mesh scale $s$ with $0.05 < s < 0.85$. When $s$ is increased more and smaller mesh elements are introduced, and a better resolved mesh results. For a given field component  $g$, we make a semi-log plot $\eps_2(g_s,g_{0.85})$ versus $s$ of the relative deviation $\eps_2(g_s,g_{0.85})$ of the solution $g_s$ with mesh scale $s$ from the solution $g_{0.85}$ with the maximum mesh scale $s=0.85$ (the finest mesh). For all fields (in 1st order $\pI$, $\vnn_{1,i}$ with $i = x,y(,z)$, $\unn_{1,i}$ with $i = x,y(,z)$, and $\vph_1$, and in 2nd order $\pII$, $\vnn_{2,i}$ with $i = x,y(,z)$), we find that $\eps_2(s)$ for $s\gtrsim 0.2$ is well described by an exponential decay, which indicates that convergence is obtained. In particular for $s = 0.7$, all relative deviations are small, $\eps_2(0.7) < 0.01$, and we therefore choose this value of $s$ as our converged mesh. As in Ref.~\cite{Muller2012}, the 1st-order fields converge faster than the 2nd-order fields.

The computational time and memory requirement for the various simulations using the converged mesh scale $s=0.7$ are as follows. In 2D, all simulations are done using COMSOL 6.2 on a HP-Z4 workstation with a processor Intel(R) Xeon(R) W-2295 CPU @3.00~GHz and 512 GB of random access memory (RAM). All 2D simulations uses the same 2D mesh, which has 23,145 mesh elements with 12,712 mesh vertices, which for the Full model results in 543,107~degrees of freedom (DoF), a computational time of 48 s, and a memory usage of 16.8 GB RAM, and which for the BL25 model results in 363,915~DoF, a computational time of 30 s, and a memory usage of 14.8 GB RAM.

For the 3D simulations, despite the small-sized device geometry, the computational time and memory requirement of the Full model forced us to use the DTU HPC cluster. To obtain a fair comparison of the computational time requirement between the Full model and the BL25 model, we therefore also ran the latter on the DTU HPC cluster. In 3D, the Full model mesh has $101,839$ mesh elements with $25,060$ vertices, which results in 2,536,778 DoF, a computational time of 26 min 26 s, and a memory usage of 348~GB RAM. In contrast, the 3D mesh of the BL25 model has only $35,098$ mesh elements with $6,344$ vertices, which results in 278,078 DoF, a computational time of 1 min 11 s, and a memory usage of 33~GB RAM. Here we clearly see the huge reduction in the computational requirements, when going from the Full model to the BL25 model in 3D simulations. We end by noting that the much higher computational cost of the Full model is caused not only by the fine resolution needed in the thin boundary layer, which by construction is avoided in the BL25 model, but also because to obtain the same accuracy, the Full model requires a finer mesh in the bulk than the BL25 model.

\subsection{3D numerical validation of BL25 vs.\ Full model:
the tiny PMMA device with an ellipsoidal cavity}
\seclab{numval3D}

We now turn to the numerical validation of the BL25 model by comparing it to the Full model. This is done on the tiny sub-mm PMMA-based ellipsoidal-cavity device actuated by a thin-film transducer described in \secref{device3D}. The comparison covers simulation of the 1st- and 2nd-order fields mentioned in \secref{numValidation}, and we remark that to the best of our knowledge, the Full model simulations are the first of its kind carried out for a 3D acoustofluidic device.

\begin{figure}[t]
\centering
\includegraphics[width=\columnwidth]{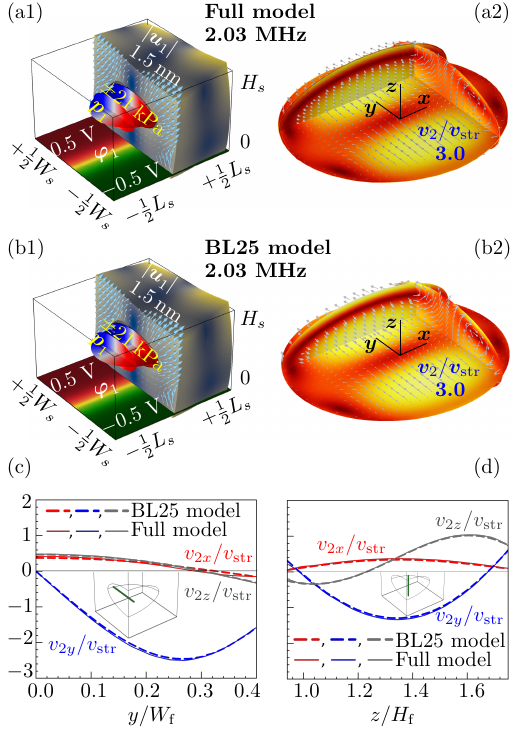}
\caption[]{\figlab{2nd-order_3D}
Validating the 1st- and 2nd-order fields of the BL25 model against the Full model for the tiny sub-mm 3D PMMA device actuated by a piezoelectric split-top-electrode thin-film transducer at the resonance frequency $\fres = 2.03$~MHz, see \figref{device_sketch}(a) and \tabref{device_dimensions}. (a1) Color plot of the Full model 1st-order pressure $p_1$ from $-21$~kPa (blue) to $+21$~kPa (red), the amplitude of the displacement field, $|\uuu_1|$ from $0$ (dark blue) to $1.5$~nm (yellow), and the electrical potential $\vphn_1$  from $-0.5$~V (green) to $+0.5$~V (red). (a2) Vector plot of the Full model streaming velocity $\vvv_2$ and color plot of its normalized magnitude $\vvn_2/v_\mr{str}$ from $0$ (Dark brown) to $3.0$. (b1) Same as (a1), but for the BL25 model. (b2) Same as (a2), but for the BL25 model. (c) Line plots of the streaming field components comparison for the Full and BL25 models along the green line parallel to the $y$-direction. (d) Same as (c), but plotting along the green line parallel to the $z$-axis.}
\end{figure}

The thin-film transducer is attached at the bottom of the device as shown in \figref{device_sketch}(a), and the device is actuated by applying an AC voltage with frequency $f$ and amplitude $+\frac12 \vphn_0$  ($-\frac12 \vphn_0$) on the top left-half (right-half) of the split top electrode, $\vphn_0 = 1~\SIV$, and by grounding the bottom electrode. Using the BL25 model, the resonance frequency for the PMMA device is obtained as the position $\fres = 2.03$~MHz of the maximum energy density $\Eac(f)$ in a frequency sweep from $0$ to $2.5$~MHz in steps of $5.0$~kHz, and we find that $\Eac(\fres) = 0.06\:\SIJpcm$. Then we compute the 1st- and 2nd-order acoustic fields in both the BL25 and the Full model at $\fres$ . The comparison of 1st- and 2nd-order fields in the two models are presented in \figref{2nd-order_3D}(a1) and \figref{2nd-order_3D}(b1), respectively. The relative deviation between the models are $\eps^\mr{1st}_2 \lesssim 0.5~\%$ and  $\eps^\mr{2nd}_2 \lesssim 2~\%$ for the 1st- and 2nd-order fields, respectively, and both models find that for example that the 1st-order acoustic pressure inside the fluid channel has the amplitude  $21~\SIkPa$ and that the maximum solid displacement of the PMMA is $1.5$~nm. The 2nd-order velocity (streaming) fields are notoriously difficult to compute. We obtained it in the BL25 model by computing $\vvv_2^d$ using \eqref{cont_navier_second_long} along with the slip boundary conditions~\eqnoref{v20Final}, and in the Full model we solved the 2nd-order \eqref{cont_navier_second} to obtain the streaming fields also inside the numerically resolved boundary layers. The contour plots of the streaming velocity $\vvvII$ for the 3D tiny PMMA device obtained in the two models are shown in \figref{2nd-order_3D}(a2) and \figref{2nd-order_3D}(b2), respectively. We obtained the same streaming pattern in the two models in the bulk of the fluid volume, including the conventional Rayleigh streaming pattern with the four streaming rolls in the $yz$-plane, as well as the streaming pattern in the central $xy$-plane and $xz$-plane. In the plots we have normalized the streaming fields by the amplitude $\vstr = \frac38\:\frac{v_1^2}{\cO}  = 0.06 \: \SImumps$ of the conventional Rayleigh streaming.

The good agreement between the BL25 and Full model is further corroborated by plotting the respective streaming velocity components along the two lines passing through the point $(0.1L_f,0.1W_f,0.5H_f)$ parallel to the $y$- and $z$-axis respectively. The resulting line graphs are plotted in \figref{2nd-order_3D}(c,d), where minor deviations can be seen closer to the boundary layers, which can be attributed to the coarser mesh used in the Full model. Since we are at the limit of the computational capability for using the Full model, the boundary layer in this model was not fully resolved, but nevertheless, the relative deviation in terms of $L^2$-norms for the line plots were $\eps_2 \leq 1.7\%$.

\subsection{2D numerical validation of BL25 vs.\ Full model:
normal-sized PMMA device with an elliptic cavity}
\seclab{numval2D}

As the final part of the numerical validation of the BL25 model, we consider the 2D cross-section of the normal-sized long, straight PMMA device with an elliptic water-filled microchannel, which was introduced in \figref{device_sketch}(b) together with the parameter lists in \tabsref{device_dimensions}{material_values}. Like conventional acoustofluidic devices, the 2D PMMA device is actuated by a bulk-sized piezoelectric transducer (PZT) glued by a thin layer of glue to the bottom surface of the PMMA block. Going to 2D has the advantage that the problems related to the extensive computer memory requirements in 3D are avoided, and we can also redo the numerical validation of the BL18 model, which was done only in 2D in the original work~\cite{Bach2018}. For better comparison, we have used the same mesh in Full and BL models in the present analysis, although a coarser mesh would suffice for the BL models~\cite{Bach2018}.

Similar to the 3D device, the transducer of the 2D has a split top electrode, and it is actuated by applying an AC voltage with frequency $f$ and amplitude $+\frac12 \vphn_0$  ($-\frac12 \vphn_0$) on the top left-half (right-half) of the split top electrode, $\vphn_0 = 1~\SIV$, and by grounding the bottom electrode. As in 3D, the resonance frequencies $\fres$ of the 2D PMMA device are obtained as the those leading to local maxima $\Eac(\fres)$ when computing the acoustic energy density $\Eac$ versus frequency from $0.5$ MHz to $2.5$ MHz in steps of $5$ kHz. The relevant resonance frequency here is $\fres = 0.98~\SIMHz$, which clearly, and in accordance with the WSUR principle for acoustically soft (PMMA) devices~\cite{Moiseyenko2019}, is much lower than the corresponding resonance mode ($\sim 2~\SIMHz$) in an acoustically hard (glass) device.

\begin{figure}[t]
\centering
\includegraphics[width=\columnwidth]{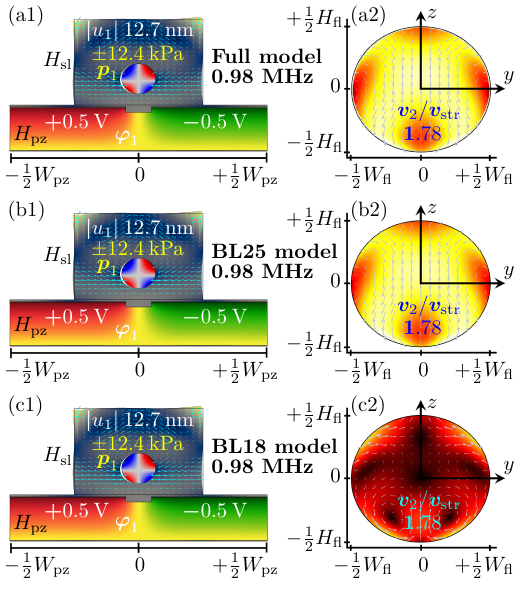}
\caption[]{\figlab{1st_2nd-order_2D}
Validating the 1st- and 2nd-order fields in the BL25 and BL18 model against the Full model for the mm-sized 2D PMMA device actuated by a piezoelectric split-top-electrode bulk transducer at the resonance frequency $\fres = 0.98$~MHz, see \figref{device_sketch}(b) and \tabref{device_dimensions}. (a1) Full model 1st-order fields:  pressure field, $p_1$ from $-12.4$~kPa (blue) to $+12.4$~kPa (red), the magnitude of solid displacement field $|\unn_1|$ from 0 (Dark blue) to 12.4~nm (yellow), the 1st-order electrical potential from $-0.5$~V (green) to $+0.5$~V (red) via the grounded electrode value of $0$~V (yellow). (a2) Full model 2nd-order streaming field normalized with the Rayleigh streaming velocity from $0$ (Dark red) to $1.78$ (yellow). (b1) Similar to (a1), but for the BL25 model. (b2) Similar to (a2), but but for the BL25 model. (c1) Similar to (a1), but for the BL18 model. (c2) Similar to (a2), but for the BL18 model.}
\end{figure}

In \figref{1st_2nd-order_2D}(a1,b1,c1) are shown the simulated contour plots of the 1st-order fields for the 2D PMMA device at resonance, $\fres = 0.98~\SIMHz$, obtained from the Full model, the BL25 model, and the BL18 model, respectively. The resonance mode here, is a mixed-mode type with two pressure nodal lines (planes in 3D) parallel to $y$ and $z$ direction. The amplitude of the 1st-order acoustic pressure $\pI$ inside the channel is found to be $12.4~\SIkPa$, and the maximum displacement $\unn_1$ of the PMMA is $12.7~\SInm$. All three models predict the same 1st-order fields, and the relative deviations from the Full model is small, $\eps_2^\mr{1st} \lesssim 0.01$ for both the BL25 and the BL18 model.

In \figref{1st_2nd-order_2D}(a2,b2,c2) are shown the simulated contour plots of the 2nd-order streaming field for the 2D PMMA device at resonance, $\fres = 0.98~\SIMHz$, obtained from the Full model, the BL25 model, and the BL18 model, respectively. Whereas the BL25 model still has a small relative deviation from the Full model, $\eps_2(v_2^\mr{BL25},v_2^\mr{FM}) = 2~\%$, the BL18 model fails completely both quantitatively with  $\eps_2(v_2^\mr{BL18},v_2^\mr{FM}) = 96~\%$ and predicting a qualitatively wrong solution. We note that when computing the $L^2$-norm we exclude the region closer than $15~\SImum$ to avoid the boundary layers where the BL-models by construction deviate from the Full model. In this mixed-mode resonance, the conventional Rayleigh streaming pattern with four flow rolls does not appear, but for convenience we still plot the streaming fields normalized with the Rayleigh streaming velocity  $\vnn_\mr{str} = \frac38 \frac{v_1^2}{\cO} = 0.06~\SImumps$. This comparison highlights the necessity of replacing the BL18 model by the BL25 model.

To further study the extend to which the BL25 and BL18 reproduce the Full model results for the 2nd-order streaming velocity, we supplement the current elliptical channel cross section with eccentricity 0.44 by another with eccentricity 0.87, both embedded in PMMA (soft) and in glass (hard). The respective relative deviations $\eps_2$ from the Full model are listed in \tabref{L2norm_calc}.

\begin{table}[t]
\centering
{
\caption{\tablab{L2norm_calc} The relative deviation $\eps_2(v_2^\mr{B},v_2^\mr{F})$ of the BL25 and BL18 models from the Full model simulation of the acoustic streaming $\vvv_2$ for elliptic channels with eccentricity $e=0.44$ [\figref{device_sketch}(b)] and $e=0.87$, embedded in (soft) PMMA and in (hard) glass, and actuated at the frequency $f$. Boldfaced entries are computed at resonance.}
\begin{ruledtabular}
\begin{tabular}{c@{\qquad}cc@{\qquad}cc}
  $f$ &
  \multicolumn{2}{l}{2D device with PMMA} &
  \multicolumn{2}{l}{2D device with glass} \\
       (MHz)      &
  \makebox[15mm][c]{$\eps_2(v_2^\mr{25},v_2^\mr{F})$}  &
  \makebox[15mm][c]{$\eps_2(v_2^\mr{18},v_2^\mr{F})$}  &
  \makebox[15mm][c]{$\eps_2(v_2^\mr{25},v_2^\mr{F})$}  &
  \makebox[15mm][c]{$\eps_2(v_2^\mr{18},v_2^\mr{F})$}  \\  \hline
  \multicolumn{5}{l}{\textit{Elliptical channel cross section with eccentricity $e = 0.44$ \upspace}}\\
 $0.98$   &  \textbf{4~\%} & \textbf{96~\%} &  3~\% & 16~\%\upspace \\
 $1.52$   &  \textbf{2~\%} & \textbf{13~\%} &  1~\% & 27~\%  \\
 $3.66$   &  3~\% & 24~\% & \textbf{1~\%} &  \textbf{7~\%}  \\
 $4.02$   &  1~\% & 14~\% & \textbf{1~\%} &  \textbf{9~\%}  \\
   \multicolumn{5}{l}{\textit{Elliptical channel cross section with eccentricity $e = 0.87$ \upspace}}\\
 $0.98$   &  \textbf{3~\%} & \textbf{148~\%} &  3~\% & 20~\%\upspace  \\
 $1.97$   &  1~\% & 25~\% &  \textbf{1~\%} &  \textbf{2~\%}  \\
 $3.95$   &  \textbf{1~\%} & \textbf{13~\%} &  2~\% & 36~\%  \\
 $4.27$   &  1~\% & 18~\% &  \textbf{2~\%} & \textbf{16~\%}  \\
\end{tabular}
\end{ruledtabular}
}
\end{table}

Clearly, the BL25 model is more accurate than the BL18 model. The relative deviation of the BL25 model from the Full model is at most 4~\%, whereas the BL18 model deviates typically 10~\% or more. However, we note that the BL18 model is performing better for the hard glass devices than the soft PMMA devices, and the best performance of the BL18 model is at resonance frequencies in hard devices, exactly the case that have been tested previously~\cite{Bach2018, Skov2019}. The reason for this behavior is that the pre-factors of the terms in the higher orders of $k\delta$ previously neglected, are so large that these term cannot be neglected. The inclusion of these term is exactly what distinguishes the BL25 model from the BL18 model, and we now see the result: The BL25 model is accurate both on and off resonance for both hard and soft devices. It is worth mentioning that the computation time of the complex and more accurate BL25 model is not significantly different from the more simple BL18 model. The reason is that the model complexity only affects the short time for initializing the finite-element matrix, but not the subsequent long time spend by the finite-element solver.

So far, we have only discussed geometries without sharp corners. However, the right angles in a rectangular channel cross section are sharp with a zero curvature. Thus, this common geometry breaks the fundamental assumption of weak curvature, on which the boundary-layer models are founded. To investigate corner effects, we have simulated 2D rectangular devices, and the results are presented in the Supplementary Material~\cite{Note1}. In summary, we found that in (hard) glass devices on resonance, the relative deviations are small for both BL25 and BL18, $\eps_2(v_2^\mr{25},v_2^\mr{F}) = 0.4~\%$ and $\eps_2(v_2^\mr{18},v_2^\mr{F}) = 2~\%$, but for the (soft) PMMA device, only BL25 remains accurate whereas BL18 fails, $\eps_2(v_2^\mr{25},v_2^\mr{F}) = 4~\%$ and $\eps_2(v_2^\mr{18},v_2^\mr{F}) = 78~\%$. So, also regarding corner effects in rectangular channels, the replacement of the BL18 model by the BL25 model is  necessary for accurate simulations.

\begin{figure*}[t]
\centering
\includegraphics[width=0.9\textwidth]{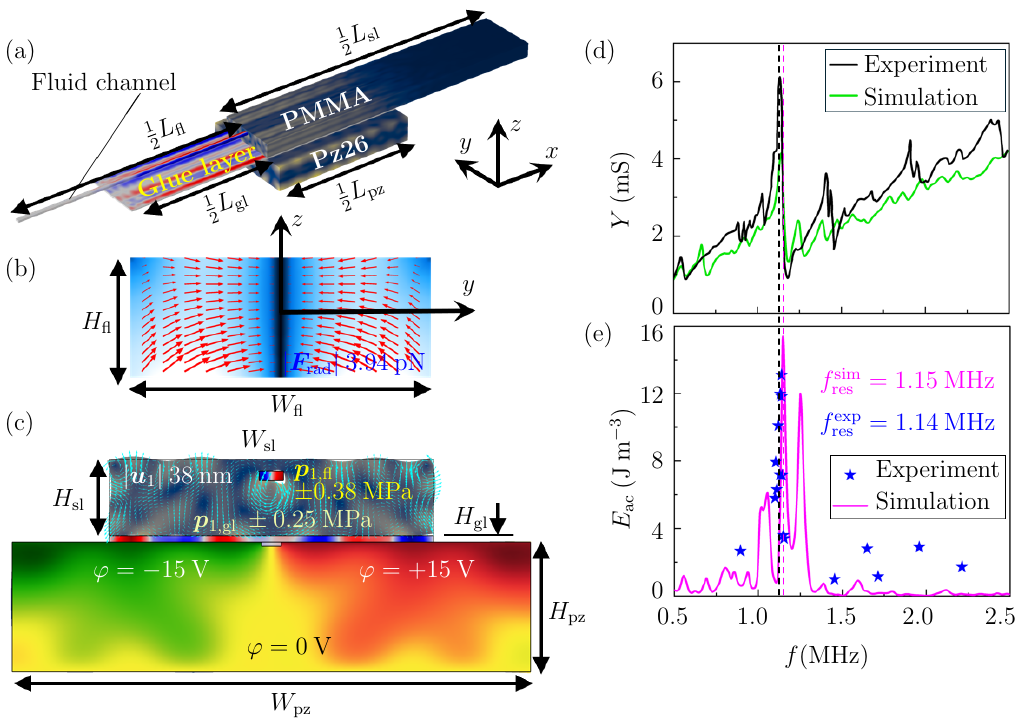}
\caption[]{\figlab{exp-sim3D}
BL25-model simulations in 3D of the PMMA device studied by Lickert \etal~\cite{Lickert2021} with the parameters listed in~\tabssref{material_values}{Lickert_device_dimensions}{Pz26_param}. (a) Simulated 1st-order fields in 3D at the resonance frequency $\fres =1.15$~MHz: the pressure inside the fluid channel [$p_{1,\fl}$ from $-0.38$~MPa (blue) to $+0.38$~MPa (red)] and inside the glue layer [$p_{1,\mr{gl}}$ from $-0.25$~MPa (blue) to $+0.25$~MPa (red)]. (b) A vector plot (red) of the simulated acoustic radiation force $\FFFrad$ and a color plot of its magnitude $|\Frad|$ from 0 (black) to $3.94$~pN (white) in the $yz$ plane at $x=0$. (c) Color and vector plots of the simulated 1st-order fields $\pI$ (both channel and coupling layer), $\uun_1$, and $\vphn_1$ in the cross-section in the $yz$-plane at $x=0$. (d) The measured (black) and the simulated (green) electrical admittance $Y$ plotted versus frequency $f$ exhibiting a strong resonance near $f = 1.15~\SIMHz$. (e) Measured (crosses) and simulated (line) acoustic energy density $\Eac$ plotted versus frequency $f$. The measured and simulated resonance frequency $\fres$ and corresponding $\Eac(\fres)$ are found to be $\fres^\mr{exp} = 1.14$~MHz with $\Eac^\mr{exp} = 13\:\SIJpcm$ and $\fres^\mr{sim} = 1.15$~MHz with $\Eac^\mr{sim} = 15\:\SIJpcm$, respectively}.
\end{figure*}

%
%

\section{Model comparison with experimental data}
\seclab{expValidation}
Lastly, for an experimental validation of the BL25 model, we revisit the PMMA device (see \figref{exp-sim3D}(a)) fabricated and tested by Lickert \etal~\cite{Lickert2021} to demonstrate by experiments and by numerical BL18-simulations that (soft) polymer-based acoustofluidic devices can perform acoustophoretic particle focusing. We show below that using the BL25 model together with updated values of the acoustic parameters of PMMA, we are able to predict quantitatively the measured resonance frequency $\fres$ within 1~\%, and both the electrical admittance $Y$ and the acoustic energy density $\Eac$ within 15~\%. These results are an improvement of the BL18-model simulations of Ref.~\cite{Lickert2021}, which only predicted the measured $\fres$ with quantitative accuracy, while $Y$ and $\Eac$ were predicted merely qualitatively.

\begin{table}[t]
\centering
{
\caption{\tablab{Lickert_device_dimensions} The geometry parameters (Parm.) and their values for the PMMA-based acoustofluidic device of Ref~\cite{Lickert2021} used in the present BL25-model simulation of the device.}
\begin{ruledtabular}
\begin{tabular}{cr@{\qquad\quad}cr@{\qquad\quad}cr}
Parm.  & Value  &  Parm. & Value &  Parm. & Value \\ \hline
$L_\pz$ & $24~\SImm$ & $W_\pz$ & $8~\SImm$ & $H_\pz$ & \upspace $2~\SImm$ \\
$L_\sl$ & $50~\SImm$ & $W_\sl$ & $5~\SImm$ & $H_\sl$ &  $1.18~\SImm$ \\
        & & $H_\mr{gl}$ & $100~\SImum$& $H_\sl^\mr{lid}$ &  $0.18~\SImm$ \\
$L_\gr$ & $24~\SImm$ & $W_\gr$ & $300~\SImum$ & $H_\gr$  & $65~\SImum$ \\
$L_\fl$ & $40~\SImm$ & $W_\fl$ & $375~\SImum$ & $H_\fl$  & $150~\SImum$ \\
\end{tabular}
\end{ruledtabular}
}
\end{table}

The current experimental validation of the BL25 model was carried out in a seven-step procedure: (1) The acoustic parameters of the PMMA and the PZT transducer model Pz26 from CTS Ferroperm~\cite{CTS_Ferroperm_Pz27} were determined using our in-house ultrasound electrical induced spectroscopy (UEIS) method~\cite{Bode2022}. This contrasts the use of average literature values used in Ref.~\cite{Lickert2021}. (2) The fluid properties for the iodixanol-water~16~\% V/V and glycerol-water~99~\% V/V solutions are set to the table values also used by Lickert \etal~\cite{Lickert2021}. (3) The ill-determined thickness of the 'glue' layer (here, the 'glue' is actually the very viscous glycerol-water solution, but for simplicity still called 'glue') was initially set to $20~\SImum$ as in Ref.~\cite{Lickert2021}. (4) The BL25 model was implemented in the water-filled channel where $\delta \approx 0.5~\SImum$, whereas the Full model was implemented in the highly viscous glycerol 'glue' layer, where $\delta \approx 20~\SImum$ is so large that resolving the boundary layer numerically is unproblematic. (5) The simulation was then executed.  The solid and fluid domain was discretized with tetrahedral mesh elements of maximum size, $h_{\text{solid}}^{\text{max}}= 375 \: \SImum$ and $h_{\text{fluid}}^{\text{max}}= 93.75 \: \SImum$, respectively, resulting in a mesh with 88,313 mesh elements, 13,670 boundary elements, and 1,711 edge elements, and the total DoF being 1,546,206. (6) We found that the glue-layer thickness plays a crucial role in damping the acoustic energy density in the fluid channel. Glycerol was used in the experiment to allow for easy mounting and dismounting of the device, but uniformity of the 'glue' layer thickness is difficult to control experimentally. Its thickness was not measured in the experiment, but merely assumed to lie in the range from 20 to $100~\SImum$. A study of the effects of the glue-layer thickness on the acoustic energy density of the fluid is presented in the Supplemental Material~\cite{Note1}. There we conclude that a thickness of $H_\mr{gl} = 100~\SImum$ best fitted the experimental value of $Y$. (7) With $H_\mr{gl} = 100~\SImum$, we ran the BL25-model as a function of frequency from 0.5 to $2.5~\SIMHz$ the same range as in Ref.~\cite{Lickert2021}.

The simulation results are shown in \figref{exp-sim3D} including the 1st-order fields (pressure $\pI$, displacement $\uuuI$, and electric potential $\vphn_1$), the acoustic radiation force~\eqref{FradDef} $\FFFrad$, and notably, the averaged acoustic energy density~\eqref{EacDef} $\Eac$ in the fluid, and the electrical admittance spectrum $Y(f)$ of the transducer loaded with PMMA and the fluid channel. $Y$ is computed as the current $I$ flowing into the positive electrode surface $\Omega_+$ over the voltage difference $\vphn_0$ between the positive and the negative electrode, which in terms of $D_z$ of \eqref{StressStrainPiezo} is,
 \beq{YDef}
 Y(f) = \frac{1}{\vphn_0} \int_{\Omega_+} \JJJ\cdot(-\nnn)\:\dm a = \frac{\ii\omega}{\vphn_0}\int_{\Omega_+} D_z\:\dm a.
 \eeq
The electrical admittance spectrum is presented in \figref{exp-sim3D}(d). In the BL25-model simulation, the maximum admittance value $Y_\mr{max}^\mr{sim}=4.4~\SImS$ is obtained at the frequency $\fres^{\mr{sim}} = 1.15~\SIMHz$, which defines the resonance frequency. This is in fair agreement with the measured values $Y_\mr{max}^\mr{exp}=6.0~\SImS$ obtained at $\fres^{\mr{exp}} = 1.13~\SIMHz$. Also off-resonance in the entire frequency range, the BL25-model predicts fairly well within 10 - 30~\% the entire admittance spectrum.

These results are consistent with the acoustic energy spectrum, presented in \figref{exp-sim3D}(e), where in the BL25 model, the simulated maximum energy value $E_\mr{ac,max}^\mr{sim}=15~\SIJpcm$ is obtained at $\fres^{\mr{sim}} = 1.15~\SIMHz$, the same frequency as for $Y_\mr{max}^\mr{sim}$, and the measured maximum energy value $E_\mr{ac,max}^\mr{exp}=13~\SIJpcm$ is obtained at $\fres^{\mr{sim}} = 1.13~\SIMHz$, the same frequency as for $Y_\mr{max}^\mr{exp}$. Thus $E_\mr{ac,max}^\mr{sim}$ deviates only 15~\% from $E_\mr{ac,max}^\mr{exp}$, which an improvement of the numerical model compared to the
550~\% deviation found in Ref.~\cite{Lickert2021}. In short, we find that the BL25 model is validated against experimental data from the literature. The predicted resonance frequency deviates only 1~\% from the measured one, and the predicted admittance and acoustic energy density deviates around 15~\%.

We end this section by noting that we have observed that in the BL25 model the above fluid resonance frequency does not deviate from the resonance frequency of the unloaded Pz26 transducer, which in simulation and experiment was found to be $\fres^{\pz, \text{sim}} = 1.145$~MH and $\fres^{\pz, \text{exp}} = 1.14$~MHz, respectively, in contrast to the miss-match between the resonance frequency of the unloaded transducer and the fluid reported by Lickert \etal~\cite{Lickert2021}.

The BL25 model corroborates the ability of the soft PMMA device to perform particle focussing. In \figref{exp-sim3D}(c) is shown, how for an applied voltage of $30~\SIV_\mr{pp}$, a strong displacement ($28~\SInm$) of the PMMA occurs just below the fluid channel in a pattern which creates an antisymmetric wave that leads to the formation of half standing wave inside the fluid channel with the pressure node aligned vertically along the fluid center of magnitude $0.38~\SIMPa$. For application purposes, the more interesting quantity to study, is the acoustic radiation force acting on suspended microparticles in the fluid channel, as shown in \figref{exp-sim3D}(b) for 5-$\SImum$-diameter polystyrene particles as used in Ref.~\cite{Lickert2021}. The maximum force obtained from the simulation is $3.9~\SIpN$ in the bottom half of the microchannel, and it is seen how the horizontal component generally is larger than the vertical one. This is quantified by the figure of merit~\cite{Moiseyenko2019}, defined as  $R = \int_{V_\mr{fl}} -\mr{sign}(y) \Frad_y\:\dm V/\int_{V_\mr{fl}} \big|\Frad_z\big|\:\dm V = 3.1$, a value comparable to the ones found in the ideal 2D cases studied in Ref.~\cite{Moiseyenko2019}, and indicating a acoustophoretic ability of decent quality.

%
%

\section{Summary and concluding remarks}
\seclab{conclusion}
The core of the theory developed by Bach and Bruus~\cite{Bach2018} is the decomposition of the 1st- and 2nd-order acoustic fields into short-range boundary-layer fields, which vary on the length scale  $\delta$, and long-range bulk fields, which vary on the length scale of the inverse of wave number $k_0^{-1}$. This decomposition enables an analytical evaluation of the boundary layer field given the assumptions that the wall motion is small and the fluid-solid curvature is large. With these prerequisites, Bach and Bruus developed the BL18 model in terms of a slip-velocity boundary condition for the streaming velocity field valid to lowest order in the small parameter $\kO\delta \approx 0.003$. Since the BL18 model avoids to resolve the thin boundary layer numerically, it enables simulation of actual acoustofluidic devices in 3D. So far, the BL18 model has only been tested for (hard) glass and glass-Si devices~\cite{Bach2018, Skov2019}, but remarkably, we have demonstrated in this work, that for soft polymer-based devices, the higher order $k_0\delta$ terms, hitherto neglected in the BL18 model, must be included to obtain a quantitative correct description of (soft) polymer acoustofluidic devices, even in cases where the BL18 model fails to predict qualitative correct responses. We call this improved model the BL25 model.

In \secref{BC_p1}, the boundary condition derived for the 1st-order pressure $\pI$ in the BL25-model is identical to that of the BL18 model. However, in \secref{BC_sigma1} one extra terms is added to the 1st-order stress-boundary condition in the BL25 model compared to the BL18, the first difference between the two models.

Going to the 2nd-order streaming field $\vvn_2$, the two models differ more substantially. We found that for acoustically soft materials like polymers, it is important to include terms of one higher order in $k\delta$ than was done in the BL18 model, as the polymer wall velocity can be more than one order of magnitude larger than that of a hard glass or glass-Si device. In particular, the larger wall velocity facilitates a larger short-ranged perpendicular boundary-layer velocity component $\vds_{2\perp}$, resulting in the non-negligible influence of the higher-order terms in the short-range boundary-layer velocity $\vvvds_2$ \eqref{vddFinal}.

The main theoretical result of this work is the final form of the slip boundary condition~\eqref{v20Final} of the BL25 model, comprising a total of 25 terms for the long-range streaming field for $\vvvls_2$, obtained by combining short-range velocity  $\vvvds_2$ and the Stokes drift velocity $\vvv^\mr{sd}_2$, \eqsref{vddFinal}{StokesDriftWall}.

The main modeling result is the implementation of the BL25 model in the finite-element method using the \qmtt{Weak Form PDE Interface} of the software COMSOL Multiphysics, described in \secref{COMSOL_impl}. Here, the detailed implementation of the tangential and perpendicular derivatives on the weakly-curved fluid-solid interface, which appears in the above mentioned boundary conditions, is essential.

Finally, the main simulation results are the numerical and experimental validation of the BL25 model presented in \secsref{numValidation}{expValidation} based on the 3D and 2D models introduced in \secref{device_model} and the experimental device described in \secref{expValidation}, all shown in \figsref{device_sketch}{exp-sim3D}, respectively.
The first step in the numerical validation was the mesh convergence study presented in \secref{MeshConv2D3D}, which revealed satisfactory convergence of the 1st- and 2nd-order fields in the BL25 model. In all cases, an asymptotic behavior approached exponentially was obtained for all fields as a function of refining the mesh.

The second numerical validation step was the comparison in 3D between the BL25 model and the Full model presented in \secref{numval3D}. This validation is not trivial as the main point in developing the BL25 model is to be able to perform 3D device simulation. However, by introducing a tiny sub-mm model device, still containing all the essential components of an acoustofluidic device, we were able to simulate the acoustofluidic response in both models by stretching our available computer resources to the limit. To our knowledge, this may be the first example of a Full-model simulation of an acoustofluidic device. This validation was successful, as we found the relative deviation between the models to be $\eps^\mr{1st}_2 \lesssim 0.5~\%$ and  $\eps^\mr{2nd}_2 \lesssim 2~\%$ for the 1st- and 2nd-order fields, respectively.

The third numerical validation step, presented in \secref{numval2D}, was to compare simulation results in 2D of both the BL25 model and the BL18 model to the Full model. The results are summarized in \figref{1st_2nd-order_2D} and \tabref{L2norm_calc}. The BL25 model predicts the results of the Full model with both qualitative and quantitative accuracy, with relative deviations from the Full model being at most 4~\% in all shown cases, including both hard glass and soft PMMA devices being driven both on and off resonance. In contrast, the BL18 model only obtains the same good level of quantitative agreement with the Full model in the cases previously tested, hard devices run at resonance~\cite{Bach2018, Skov2019}, whereas large quantitative deviations are seen for soft polymer devices both on and off resonance, and somewhat surprising, also for hard glass devices run off resonance. Importantly, we even found cases where the predictions of the BL25 model matched the Full model, whereas the BL18 model predicted qualitative wrong responses, such as the example shown in \figref{1st_2nd-order_2D}(c2). In a final 2D numerical validation, we found that even in the case of devices with rectangular channel cross sections, where the four corners with their zero radius of curvature clearly violate the basic assumption of the BL25 and BL18 models that the curvature must be sufficiently large compared to the boundary-layer thickness $\delta$, the BL25 model deviated very little from the Full model, whereas the BL18 model failed. Specifically, we found the relative deviations $\eps_2(v_2^\mr{25},v_2^\mr{F}) = 4~\%$ and $\eps_2(v_2^\mr{18},v_2^\mr{F}) = 78~\%$ in the (soft) PMMA device run at resonance. For a (hard) glass device at resonance, both models did better, with $\eps_2(v_2^\mr{25},v_2^\mr{F}) = 0.4~\%$ and $\eps_2(v_2^\mr{18},v_2^\mr{F}) = 2~\%$.

Finally, the experimental validation against published data in the literature by Lickert \etal~\cite{Lickert2021} for a (soft) PMMA device driven by a PZT Pz26 transducer presented in \secref{expValidation}, show that the BL25 model can predict the measured resonance frequency within 1~\%  and the measured electrical admittance of the loaded PZT Pz26 transducer as well as the acoustic energy density in the water-filled channel  both qualitatively and quantitatively (within about 15~\%). In this validation, the thickness of the glycerol 'glue' layer was the only fitting parameter. This is in contrast to the previous BL18 model results also published in Ref.~\cite{Lickert2021}, where the electrical admittance and the acoustic energy density deviated by nearly an order of magnitude from the experimental data. This experimental validation could be improved by knowing the specific clamping of the device in the sample holder and the actual thickness of the 'glue' layer.

Given the extensive validation simulations showing that the BL25 model is in good quantitative agreement (within 4~\%) with the full model for both 1st- and 2nd-order fields and both on and off resonance, given the quantitative agreement (within 15~\%) with experimental data, and given that the BL25 model in all cases performs better than the previous BL18 model, it appears to be justified to introduce the complex BL25 model comprising of the 25 terms in \eqref{v20Final}. It is worth mentioning that the computation time of the complex BL25 model is not significantly different from the more simple BL18 model. Based on the presented results, we believe that the BL25 model will be an important tool in future modeling of actual 3D acoustofluidic devices, both as a mean to unravel the intricate physical nature of such devices, and as part of design optimization for technological applications of such devices.

\section*{Acknowledgements}
This work was supported by the \textit{ACOUSOME} project funded by the European Innovation
Council (EIC),  \textit{HORIZON EIC 2022 TRANSITION}, grant No.~101099787.

\appendix

%
%

\begin{table}[b]
\centering
\caption{\tablab{Pz26_param} Parameters of the PZT transducer of type Pz26~\cite{CTS_Ferroperm_Pz26} used in the numerical simulations of the experimental PMMA device \cite{Lickert2021} described in \secref{expValidation}.}
\begin{ruledtabular}
\begin{tabular}{lcrl}
Parameter &  Symbol  & Value & Unit
\\
\hline
Mass density\upspace & $\rhopz$ & $7700$ & $\SIkgpcm$ \\
Elastic modulus & $C_{11}$ & $168 - \ii 3.36$ & $\SIGPa$ \\
Elastic modulus & $C_{12}$ & $110 - \ii 2.20$ & $\SIGPa$ \\
Elastic modulus & $C_{13}$ & $99.9 - \ii 2.00$ & $\SIGPa$ \\
Elastic modulus & $C_{33}$ & $123 - \ii 2.46$ & $\SIGPa$ \\
Elastic modulus & $C_{44}$ & $30.1 - \ii 0.60$ & $\SIGPa$ \\
Coupling constant & $e_{15}$ & $9.86-\ii 0.20$ & $\SIC\: \SIm^{-2}$ \\
Coupling constant & $e_{31}$ & $-2.8 + \ii 0.06$ & $\SIC\: \SIm^{-2}$ \\
Coupling constant & $e_{33}$ & $14.7 - \ii 0.29$ & $\SIC\: \SIm^{-2}$ \\
Electric permittivity & $\ve_{11}$ & $828(1 - \ii 0.02)\;\epsO$ & $\SIF\: \SIm^{-1}$ \\
Electric permittivity & $\ve_{33}$ & $700(1 - \ii 0.02)\;\epsO$ & $\SIF\: \SIm^{-1}$
\end{tabular}
\end{ruledtabular}
\end{table}

\section{The constitutive equations for the piezoelectric transducers}
\seclab{constitutiveEqs}

The constitutive equations for the three types of piezoelectric transducers used in this work all have the same form. In the Voigt notation, the six independent components $\sigmapz_{1ik}$ of the symmetric stress tensor $\sigmabfpz_1$ and $\spz_{1ik} = \frac12 (\pp_i \unn_{1k} + \pp_k \unn_{1i})$ of the symmetric strain tensor $\ssspz_1$ are represented by the six-dimensional Voigt vectors $\sigmabf^\mr{V}_1$ and $\sss^\mr{V}_1$, whose transposed forms are

 \bsubal{VoigtStressStrain}
 \eqlab{VoigtStress}
 \big[\sigmabf^\mr{V}_1\big]^\textsf{T} &= \big\{
 \sigmapz_{1xx}, \sigmapz_{1yy}, \sigmapz_{1zz}, \sigmapz_{1yz}, \sigmapz_{1xz}, \sigmapz_{1xy} \big\},
 \\
 \eqlab{VoigtStrain}
 \big[\sss^\mr{V}_1\big]^\textsf{T} &= \big\{
 \spz_{1xx}, \spz_{1yy}, \spz_{1zz}, 2\spz_{1yz}, 2\spz_{1xz}, 2\spz_{1xy} \big\}.
 \esubal
These mechanical fields are supplemented by the electrical potential $\vphn_1$ and its gradient, the electric field $\EEn_1 = -\grad\vphn_1$, as well as the electric displacement field $\DDn_1$. The Voigt representation of the constitutive equations for the piezoelectric transducers in a $9\times 9$-matrix equation relating $\sigmabf^\mr{V}_1$ and $\DDn_1$ to $\sss^\mr{V}_1$ and $-\grad\vphn_1$,
 \bal
 \eqlab{StressStrainPiezo}
 &\left(\begin{array}{c} \sigmabf^\mr{V}_1\ \\[2mm] \DDn_1 \end{array}\right) = \MMM \cdot
 \left(\begin{array}{c} \sss^\mr{V}_1\ \\[2mm] -\grad \vphn_1 \end{array} \right),
 \\ \nn
 &\MMM \!=\!
 \left(\!\!  \begin{array}{ccc|ccc|ccc}
 \Cnn_{11} & \Cnn_{12} & \Cnn_{13} & 0 & 0 & 0 & 0 & 0 & \!-e^{{}}_{31}\downspace  \\
 \Cnn_{12} & \Cnn_{11} & \Cnn_{13} & 0 & 0 & 0 & 0 & 0 & \!-e^{{}}_{31}\downspace  \\
 \Cnn_{13} & \Cnn_{13} & \Cnn_{33} & 0 & 0 & 0 & 0 & 0 & \!-e^{{}}_{33}\downspace \\ \hline
 0\upspace & 0 & 0 & \Cnn_{44} & 0 & 0 & 0 &  -e^{{}}_{15} & 0\downspace  \\
 0 & 0 & 0 & 0 & \Cnn_{44} & 0 & -e^{{}}_{15} & 0 & 0\downspace  \\
 0 & 0 & 0 & 0 & 0 & \Cnn_{66}\!  & 0 & 0 & 0 \downspace \\ \hline
 0\upspace  & 0 & 0 & 0 &  e^{{}}_{15} & 0 & \ve^{{}}_{11} &  0 & 0\downspace  \\
 0 & 0 & 0 &  e^{{}}_{15} & 0 & 0 & 0 & \ve^{{}}_{11} & 0\downspace  \\
 e^{{}}_{31} & e^{{}}_{31} & e^{{}}_{33} & 0 & 0 & 0  & 0 & 0 & \ve^{{}}_{33}
 \end{array} \!\!  \right)\!.
 \eal
The parameter values of the piezoelectric transducers used in this work are given in \tabref{material_values} for \AlScN\ and PZT Pz27, and in \tabref{Pz26_param} for PZT Pz26.

%
%

\section{2nd-order boundary-layer fields}
\seclab{BC2ndOrderAppendix}

In this Appendix, we compute the 2nd-order, time-averaged, boundary-layer fields pressure $\pd_2$ and velocity $\vvvd_2$ in the fluid, based on the 1st- and 2nd-order expressions given in \secsref{fl_first_order}{fl_second_order}. Following \eqref{v2dSplit}, the velocity field is split in two components, $\vvvds_2 = \vvvddp + \vvvddv$, that are treated independently.

\subsection{Computation of $\textbf{\emph p}_2^{\delta}$}
\seclab{p2Appendix}

To obtain the coordinate-separated form of the source term $\grad \pd_2$ in \eqref{Laplvd2Final}, we first take the divergence of \eqref{Laplvd2}, neglect the $\div\vvvd_2$-terms of order $\Gamma$, and finally use $\Lapl \pd_2 \approx \ppsqr_\zs \pd_2$, to arrive at
 \beq{p2d_laplace} 
 \ppsqr_\zs \pd_2 = -\rhoO \div \big[\div \avr{\vvvd_1 \vvvd_1+ \vvvl_1 \vvvd_1+ \vvvd_1 \vvvl_1} \big].
 \eeq
Then the coordinate-separated form of this expression is obtained by using \eqsref{double_div_AB}{vd1vl1BL}, and we arrive at the following 20-term expression for $\ppperpsqr p_2^\delta$,
 \bal
 \eqlab{lapl_zeta_p2d} 
 &\ppperpsqr p_2^\delta=  -\rhoO\Big\{
 \\ \nn
 &
 +\pardiv\big[\pardiv \timeav{\intnab 0qq \vvvdspar}{\vvvdspar}\big] 
 +\timeav{\intnab {-2}qq \vdsperp}{\vdsCperp}  
 \\ \nn
 &
 +\pardiv \big[\timeav{\intnab {-1}qq \vvvdspar}{\vdsperp} 
 +\timeav{\intnab {-1}qq \vdsperp}{\vvvdspar}\big] 
 \\ \nn
 &
 +\pardiv \big[\pardiv \timeav{\intnab 01q \vvvlspar}{\vvvdspar} \big]  
 +\intnab {-2}1q  \big(\vlsperp \vdsperp \big) 
 \\ \nn
 &
 +\pardiv \big[\timeav{\intnab {-1}1q \vvvlspar}{\vdsperp} 
 +\timeav{\intnab {-1}1q \vlsperp}{\vvvdspar}\big] 
 \\ \nn
 &
 +\pardiv\big[\pardiv \timeav{\intnab 0{\zs}q \ppperp\vvvlspar}{\vvvdspar}\big] 
 +\timeav{\intnab  {-2}{\zs}q \ppperp \vlsperp}{\vdsperp} 
 \\ \nn
 &
 +\pardiv\big[\timeav{\intnab {-1}{\zs}q \ppperp\vvvlspar}{\vdsperp} 
 +\timeav{\intnab {-1}{\zs}q \ppperp \vlsperp}{\vvvdspar}\big] 
 \\ \nn\
 &
 +\pardiv \big[\pardiv \timeav{\intnab 0q1  \vvvdspar}{\vvvlspar}\big] 
 +\timeav{\intnab {-2}q1 \vdsperp}{\vlsperp} 
 \\ \nn
 &
 +\pardiv \big[\timeav{\intnab {-1}q1 \vvvdspar}{\vlsperp} 
 +\timeav{\intnab {-1}q1 \vdsperp}{\vvvlspar}\big] 
 \\ \nn
 &
 +\pardiv \big[\pardiv \timeav{\intnab 0q{\zs} \vvvdspar}{\ppperp \vvvlspar}\big] 
 +\timeav{\intnab {-2}q{\zs} \vdsperp}{\ppperp\vlsperp} 
 \\ \nn
 &
 +\pardiv \big[\timeav{\intnab {-1}q{\zs} \vvvdspar}{\ppperp \vlsperp} 
 +\timeav{\intnab {-1}q{\zs} \vdsperp}{\ppperp \vvvlspar} \big]  
 \Big\}.
 \eal
The short-range pressure $\pd_2$ is obtained from $\ppsqr_\zs\pd_2$ by integrating twice with respect to $\zeta$. This amounts to the trivial substitution of $\intnab nab$ in \eqref{lapl_zeta_p2d} by $\intnab {n+2}ab$,

\bal
 \eqlab{p2delta} 
 &p_2^\delta =  -\rhoO\Big\{
 \\ \nn
 &
 +\pardiv\big[\pardiv \timeav{\intnab 2qq \vvvdspar}{\vvvdspar}\big] 
 +\timeav{\intnab {0}qq \vdsperp}{\vdsperp}  
 \\ \nn
 &
 +\pardiv \big[\timeav{\intnab 1qq \vvvdspar}{\vdsperp} 
 +\timeav{\intnab 1qq \vdsperp}{\vvvdspar}\big] 
 \\ \nn
 &
 +\pardiv \big[\pardiv \timeav{\intnab 21q \vvvlspar}{\vvvdspar} \big]  
 +\intnab {0}1q  \big(\vlsperp \vdsperp \big) 
 \\ \nn
 &
 +\pardiv \big[\timeav{\intnab 11q \vvvlspar}{\vdsperp} 
 +\timeav{\intnab 11q \vlsperp}{\vvvdspar}\big] 
 \\ \nn
 &
 +\pardiv\big[\pardiv \timeav{\intnab 2{\zs}q \ppperp\vvvlspar}{\vvvdspar}\big] 
 +\timeav{\intnab  {0}{\zs}q \ppperp \vlsperp}{\vdsperp} 
 \\ \nn
 &
 +\pardiv\big[\timeav{\intnab 1{\zs}q \ppperp\vvvlspar}{\vdsperp} 
 +\timeav{\intnab 1{\zs}q \ppperp \vlsperp}{\vvvdspar}\big] 
 \\ \nn\
 &
 +\pardiv \big[\pardiv \timeav{\intnab 2q1  \vvvdspar}{\vvvlspar}\big] 
 +\timeav{\intnab {0}q1 \vdsperp}{\vlsperp} 
 \\ \nn
 &
 +\pardiv \big[\timeav{\intnab 1q1 \vvvdspar}{\vlsperp} 
 +\timeav{\intnab 1q1 \vdsperp}{\vvvlspar}\big] 
 \\ \nn
 &
 +\pardiv \big[\pardiv \timeav{\intnab 2q{\zs} \vvvdspar}{\ppperp \vvvlspar}\big] 
 +\timeav{\intnab {0}q{\zs} \vdsperp}{\ppperp\vlsperp} 
 \\ \nn
 &
 +\pardiv \big[\timeav{\intnab 1q{\zs} \vvvdspar}{\ppperp \vlsperp} 
 +\timeav{\intnab 1q{\zs} \vdsperp}{\ppperp \vvvlspar} \big] 
 \Big\}.
 \eal
\mbox{}\\[-10mm]\mbox{}

\subsection{Computation of $\textbf{\emph v}_2^{\delta p}$}
\seclab{v2deltapAppendix}

\mbox{}\\[-8mm]\mbox{}
The parallel component $\vvvddparp$ and perpendicular component $\vddperpp$ of the boundary-layer velocity $\vvvddp$  are obtained by integration of \eqref{Laplvddp},
\\[-6mm]
 \bsubalat{p2deriv}{3}
 \eqlab{vddparpGRADp2}
 \ppsqr_\zs \vvvddparp &= \pargrad \frac{\pd_2}{\etaO} &&
 \; \Rightarrow \;\; &
 \vvvddparp(\zeta) &= \pargrad \int^\zeta\!\!\int^{\zs'} \frac{\pd_2(\zs'')}{\etaO}\: \dm\zs''\:\dm\zs',
 \\
 \eqlab{vddperppGRADp2}
 \ppsqr_\zs \vddperpp &= \pp_\zs \frac{\pd_2}{\etaO} &&
 \; \Rightarrow \;\; &
 \vddperpp(\zs) &= \int^\zs \frac{\pd_2(\zs')}{\etaO} \dm \zs'.
 \esubalat
The computation of $\vvvddparp$ ($\vddperpp$) involves integration of $\pd_2$ twice (once) with respect of $\zs$,
which according to \eqref{intnabIntegral} amounts to increasing the superscript in all factors $\intnab nab$ by 2 (by 1). Inserting $\pd_2$ from \eqref{p2delta} into \eqsref{vddparpGRADp2}{vddperppGRADp2} leads to
\bal
 \eqlab{vddpFinalPar} 
 &\vvvddparp =  -\frac{1}{\nuO}\pargrad\Big\{
 \\ \nn
 &
 +\pardiv\big[\pardiv \timeav{\intnab 4qq \vvvdspar}{\vvvdspar}\big] 
 +\timeav{\intnab 2qq \vdsperp}{\vdsperp}  
 \\ \nn
 &
 +\pardiv \big[\timeav{\intnab 3qq \vvvdspar}{\vdsperp} 
 +\timeav{\intnab 3qq \vdsperp}{\vvvdspar}\big] 
 \\ \nn
 &
 +\pardiv \big[\pardiv \timeav{\intnab 41q \vvvlspar}{\vvvdspar} \big]  
 +\intnab 21q  \big(\vlsperp \vdsperp \big) 
 \\ \nn
 &
 +\pardiv \big[\timeav{\intnab 31q \vvvlspar}{\vdsperp} 
 +\timeav{\intnab 31q \vlsperp}{\vvvdspar}\big] 
 \\ \nn
 &
 +\pardiv\big[\pardiv \timeav{\intnab 4{\zs}q \ppperp\vvvlspar}{\vvvdspar}\big] 
 +\timeav{\intnab  2{\zs}q \ppperp \vlsperp}{\vdsperp} 
 \\ \nn
 &
 +\pardiv\big[\timeav{\intnab 3{\zs}q \ppperp\vvvlspar}{\vdsperp} 
 +\timeav{\intnab 3{\zs}q \ppperp \vlsperp}{\vvvdspar}\big] 
 \\ \nn\
 &
 +\pardiv \big[\pardiv \timeav{\intnab 4q1  \vvvdspar}{\vvvlspar}\big] 
 +\timeav{\intnab 2q1 \vdsperp}{\vlsperp} 
 \\ \nn
 &
 +\pardiv \big[\timeav{\intnab 3q1 \vvvdspar}{\vlsperp} 
 +\timeav{\intnab 3q1 \vdsperp}{\vvvlspar}\big] 
 \\ \nn
 &
 +\pardiv \big[\pardiv \timeav{\intnab 4q{\zs} \vvvdspar}{\ppperp \vvvlspar}\big] 
 +\timeav{\intnab 2q{\zs} \vdsperp}{\ppperp\vlsperp} 
 \\ \nn
 &
 +\pardiv \big[\timeav{\intnab 3q{\zs} \vvvdspar}{\ppperp \vlsperp} 
 +\timeav{\intnab 3q{\zs} \vdsperp}{\ppperp \vvvlspar} \big] 
 \Big\}.
 \eal
and
\bal
 \eqlab{vddpFinalPerp} 
 &\vddperpp =  -\frac{1}{\nuO}\Big\{
 \\ \nn
 &
 +\pardiv\big[\pardiv \timeav{\intnab 3qq \vvvdspar}{\vvvdspar}\big] 
 +\timeav{\intnab 1qq \vdsperp}{\vdsperp}  
 \\ \nn
 &
 +\pardiv \big[\timeav{\intnab 2qq \vvvdspar}{\vdsperp} 
 +\timeav{\intnab 2qq \vdsperp}{\vvvdspar}\big] 
 \\ \nn
 &
 +\pardiv \big[\pardiv \timeav{\intnab 31q \vvvlspar}{\vvvdspar} \big]  
 +\timeav{\intnab 11q \vlsperp}{\vdsperp} 
 \\ \nn
 &
 +\pardiv \big[\timeav{\intnab 21q \vvvlspar}{\vdsperp} 
 +\timeav{\intnab 21q \vlsperp}{\vvvdspar}\big] 
 \\ \nn
 &
 +\pardiv\big[\pardiv \timeav{\intnab 3{\zs}q \ppperp\vvvlspar}{\vvvdspar}\big] 
 +\timeav{\intnab  1{\zs}q \ppperp \vlsperp}{\vdsperp} 
 \\ \nn
 &
 +\pardiv\big[\timeav{\intnab 2{\zs}q \ppperp\vvvlspar}{\vdsperp} 
 +\timeav{\intnab 2{\zs}q \ppperp \vlsperp}{\vvvdspar}\big] 
 \\ \nn\
 &
 +\pardiv \big[\pardiv \timeav{\intnab 3q1  \vvvdspar}{\vvvlspar}\big] 
 +\timeav{\intnab 1q1 \vdsperp}{\vlsperp} 
 \\ \nn
 &
 +\pardiv \big[\timeav{\intnab 2q1 \vvvdspar}{\vlsperp} 
 +\timeav{\intnab 2q1 \vdsperp}{\vvvlspar}\big] 
 \\ \nn
 &
 +\pardiv \big[\pardiv \timeav{\intnab 3q{\zs} \vvvdspar}{\ppperp \vvvlspar}\big] 
 +\timeav{\intnab 1q{\zs} \vdsperp}{\ppperp\vlsperp} 
 \\ \nn
 &
 +\pardiv \big[\timeav{\intnab 2q{\zs} \vvvdspar}{\ppperp \vlsperp} 
 +\timeav{\intnab 2q{\zs} \vdsperp}{\ppperp \vvvlspar} \big]
 \Big\}.
 \eal

\subsection{Computation of $\textbf{\emph v}_2^{\delta v}$}
\seclab{v2deltavAppendix}
To determine $\vvvddv$, we compute the coordinate-separated form of the source term  in \eqref{Laplvd2Final} by using \eqsref{single_div_ABi}{vd1vl1BL},
 \balat{2}
 \eqlab{divvdvdSource}
 \ppsqr_\zs \vvvddv &= \frac{1}{\nuO}\div  \avr{\vvvd_1 \vvvd_1 + \vvvl_1 &&\vvvd_1 + \vvvd_1 \vvvl_1}
 \\ \nn
 = \frac{1}{\nuO} \Big\{&
 \pardiv \timeav{\intnab 0qq \vvvds_1}{\vvvdspar} 
 &&\!\! +\timeav{\intnab {-1}qq  \vvvds_1}{\vdsperp}  
 \\ \nn
 +&
 \pardiv \timeav{\intnab 01q \vvvls_1}{\vvvdspar}  
 &&\!\! +\timeav{\intnab {-1}1q  \vvvls_1}{\vdsperp}  
 \\ \nn
 +&		
 \pardiv \timeav{\intnab 0{\zs}q \ppperp \vvvls_1}{\vvvdspar} 
 &&\!\! +\timeav{\intnab  {-1}{\zs}q \ppperp  \vvvls_1}{\vdsperp} 
 \\ \nn
 +&
 \pardiv \timeav{\intnab 0q1  \vvvds_1}{\vvvlspar} 
 &&\!\! +\timeav{\intnab {-1}q1 \vvvds_1}{\vlsperp} 
 \\ \nn
 +&
 \pardiv \timeav{\intnab 0q{\zs} \vvvds_1}{\ppperp \vvvlspar} 
 &&\!\! +\timeav{\intnab {-1}q{\zs} \vvvds_1}{\ppperp\vlsperp}
 \Big\}.
 \ealat
From this, $\vvvddv$ is found by integrating twice after $\zs$, which according to \eqref{intnabIntegral} amounts to increase the superscript in all factors $\intnab nab$ by 2, which results in\\[-6mm]
\balat{2}
 \eqlab{vddvFinal}
 \vvvddv = \frac{1}{\nuO} \Big\{&
 \pardiv \timeav{\intnab 2qq \vvvds_1}{\vvvdspar} 
 && +\timeav{\intnab 1qq  \vvvds_1}{\vdsperp}  
 \\ \nn
 +&
 \pardiv \timeav{\intnab 21q \vvvls_1}{\vvvdspar}  
 && +\timeav{\intnab 11q  \vvvls_1}{\vdsperp}  
 \\ \nn
 +&		
 \pardiv \timeav{\intnab 2{\zs}q \ppperp \vvvls_1}{\vvvdspar} 
 && +\timeav{\intnab 1{\zs}q \ppperp  \vvvls_1}{\vdsperp} 
 \\ \nn
 +&
 \pardiv \timeav{\intnab 2q1  \vvvds_1}{\vvvlspar} 
 && +\timeav{\intnab 1q1 \vvvds_1}{\vlsperp} 
 \\ \nn
 +&
 \pardiv \timeav{\intnab 2q{\zs} \vvvds_1}{\ppperp \vvvlspar} 
 && +\timeav{\intnab 1q{\zs} \vvvds_1}{\ppperp\vlsperp}
 \Big\}.
 \ealat
Splitting this into the parallel and perpendicular components $\vvvddparv$ and ${\vddperpv}$, respectively, yields,
\balat{2}
 \eqlab{vddvFinalPar}
 \vvvddparv = \frac{1}{\nuO} \Big\{&
 \pardiv \timeav{\intnab 2qq \vvvdspar}{\vvvdspar} 
 && +\timeav{\intnab 1qq  \vvvdspar}{\vdsperp}  
 \\ \nn
 +&
 \pardiv \timeav{\intnab 21q \vvvlspar}{\vvvdspar}  
 && +\timeav{\intnab 11q  \vvvlspar}{\vdsperp}  
 \\ \nn
 +&		
 \pardiv \timeav{\intnab 2{\zs}q \ppperp \vvvlspar}{\vvvdspar} 
 && +\timeav{\intnab 1{\zs}q \ppperp  \vvvlspar}{\vdsperp} 
 \\ \nn
 +&
 \pardiv \timeav{\intnab 2q1  \vvvdspar}{\vvvlspar} 
 && +\timeav{\intnab 1q1 \vvvdspar}{\vlsperp} 
 \\ \nn
 +&
 \pardiv \timeav{\intnab 2q{\zs} \vvvdspar}{\ppperp \vvvlspar} 
 && +\timeav{\intnab 1q{\zs} \vvvdspar}{\ppperp\vlsperp}
 \Big\},
 \ealat
 \balat{2}
 \eqlab{vddvFinalPerp}
 \vddperpv = \frac{1}{\nuO} \Big\{&
 \pardiv \timeav{\intnab 2qq \vdsperp}{\vvvdspar} 
 && +\timeav{\intnab 1qq  \vdsperp}{\vdsperp}  
 \\ \nn
 +&
 \pardiv \timeav{\intnab 21q \vlsperp}{\vvvdspar}  
 && +\timeav{\intnab 11q  \vlsperp}{\vdsperp}  
 \\ \nn
 +&		
 \pardiv \timeav{\intnab 2{\zs}q \ppperp \vlsperp}{\vvvdspar} 
 && +\timeav{\intnab 1{\zs}q \ppperp  \vlsperp}{\vdsperp} 
 \\ \nn
 +&
 \pardiv \timeav{\intnab 2q1  \vdsperp}{\vvvlspar} 
 && +\timeav{\intnab 1q1 \vdsperp}{\vlsperp} 
 \\ \nn
 +&
 \pardiv \timeav{\intnab 2q{\zs} \vdsperp}{\ppperp \vvvlspar} 
 && +\timeav{\intnab 1q{\zs} \vdsperp}{\ppperp\vlsperp}
 \Big\}.
 \ealat
Adding \eqsref{vddpFinalPerp}{vddvFinalPerp}, gives the full $\vddperp$,
\\[-6mm]
 \balat{2}
 \eqlab{vddFinalPerpAppendix}
 \vddperp &= \vddperpp + \vddperpv = -\frac{1}{\nuO} \Big\{
 &&\\ \nn
 &
 +\pardiv\big[\pardiv \timeav{\intnab 3qq \vvvdspar}{\vvvdspar}\big] 
 &&+\pardiv \timeav{\intnab 2qq \vvvdspar}{\vdsperp} 
 \\ \nn
 &
 +\pardiv \big[\pardiv \timeav{\intnab 31q \vvvlspar}{\vvvdspar} \big]  %
 &&+\pardiv \timeav{\intnab 21q \vvvlspar}{\vdsperp} 
 \\ \nn
 &		
 +\pardiv\big[\pardiv \timeav{\intnab 3{\zs}q \ppperp\vvvlspar}{\vvvdspar}\big] 
 &&+\pardiv \timeav{\intnab 2{\zs}q \ppperp\vvvlspar}{\vdsperp} 
 \\ \nn
 &
 +\pardiv \big[\pardiv \timeav{\intnab 3q1  \vvvdspar}{\vvvlspar}\big] 
 &&+\pardiv \timeav{\intnab 2q1 \vvvdspar}{\vlsperp} 
 \\ \nn\
 &
 +\pardiv \big[\pardiv \timeav{\intnab 3q{\zs} \vvvdspar}{\ppperp \vvvlspar}\big] 
 &&+\pardiv \timeav{\intnab 2q{\zs} \vvvdspar}{\ppperp \vlsperp} 
 \Big\}.\\[-10mm] \nn
 \ealat

\subsection{The final expression for $\textbf{\emph v}_2^{\delta 0}$}
\mbox{}\\[-8mm]
Using the scalings  $\frac{2}{\omega \delta^2}\intBCnab nqq, \frac{2}{\omega \delta^2}\intBCnab n1q  \propto \delta^{n-2}$ and $\frac{2}{\omega \delta^2}\intBCnab n{\zeta}q  \propto \delta^{n-1}$ obtained from  \eqref{IabnBC}, we neglect the terms of higher order in $\delta$ and obtain,
 \bsuba{vddFinalAppendix}
 \bal
 \eqlab{vddFinalSum0A}
 \vvvds_2 &=  \vvvddspar + \vddsperp\:\een_\zs = \vvvddparpO + \vvvddparvO + \vddsperp\:\een_\zs,
 \\
 \eqlab{vddpFinalPar0} 
 \vvvddparpO &=  -\frac{2}{\omega \delta^2}\pargrad\Big\{
 \\ \nn
 &
 +\edbLM{23.5}{\pardiv\big[\pardiv \timeav{\intBCnab 4qq \vvvdspar}{\vvvdspar}\big]} 
 +\timeav{\intBCnab 2qq \vdsperp}{\vdsperp}  
 \\ \nn
 &
 +\edbLM{35.5}{\pardiv \big[\timeav{\intBCnab 3qq \vvvdspar}{\vdsperp} 
 +\timeav{\intBCnab 3qq \vdsperp}{\vvvdspar} \big]} 
 \\ \nn
 &
 +\edbLM{23.5}{\pardiv \big[\pardiv \timeav{\intBCnab 41q \vvvlspar}{\vvvdspar}\big]}   
 +\timeav{\intBCnab 21q \vlsperp}{\vdsperp } 
 \\ \nn
 &
 +\edbLM{35.5}{\pardiv \big[\timeav{\intBCnab 31q \vvvlspar}{\vdsperp} 
 +\timeav{\intBCnab 31q \vlsperp}{\vvvdspar}\big]} 
 \\ \nn
 &		
 +\edbLM{43.5}{\pardiv\big[\pardiv \timeav{\intBCnab 4{\zs}q \ppperp\vvvlspar}{\vvvdspar}\big] 
 +\timeav{\intBCnab  2{\zs}q \ppperp \vlsperp}{\vdsperp}} 
 \\ \nn
 &
 +\edbLM{40.0}{\pardiv\big[\timeav{\intBCnab 3{\zs}q \ppperp\vvvlspar}{\vdsperp} 
 +\dpst\timeav{\intBCnab 3{\zs}q \ppperp \vlsperp}{\vvvdspar}\big]}
 \\ \nn\
 &
 +\edbLM{23.5}{\pardiv \big[\pardiv \timeav{\intBCnab 4q1  \vvvdspar}{\vvvlspar}\big]} 
 +\timeav{\intBCnab 2q1 \vdsperp}{\vlsperp} 
 \\ \nn
 &
 +\edbLM{35.0}{\pardiv \big[\timeav{\intBCnab 3q1 \vvvdspar}{\vlsperp} 
 +\timeav{\intBCnab 3q1 \vdsperp}{\vvvlspar}\big]} 
 \\ \nn
 &
 +\edbLM{43.5}{\pardiv \big[\pardiv \timeav{\intBCnab 4q{\zs} \vvvdspar}{\ppperp \vvvlspar}\big] 
 +\timeav{\intBCnab 2q{\zs} \vdsperp}{\ppperp\vlsperp}} 
 \\ \nn
 &
 +
 \edbLM{40.0}{\pardiv \big[\timeav{\intBCnab 3q{\zs} \vvvdspar}{\ppperp \vlsperp} 
 +\timeav{\intBCnab 3q{\zs} \vdsperp}{\ppperp \vvvlspar}\big]}  
 \Big\},
 \\
 \eqlab{vddvFinalParO}
 \vvvddparvO &= \frac{2}{\omega \delta^2} \Big\{
 \pardiv \timeav{\intBCnab 2qq \vvvdspar}{\vvvdspar} 
 +\timeav{\intBCnab 1qq  \vvvdspar}{\vdsperp}  
 \\ \nn
 &+
 \pardiv \timeav{\intBCnab 21q \vvvlspar}{\vvvdspar}  
 +\timeav{\intBCnab 11q  \vvvlspar}{\vdsperp}  
 \\ \nn
 &+		
 \edbLM{20.0}{\pardiv \timeav{\intBCnab 2{\zs}q \ppperp \vvvlspar}{\vvvdspar}} 
 +\timeav{\intBCnab 1{\zs}q \ppperp  \vvvlspar}{\vdsperp} 
 \\ \nn
 &+
 \pardiv \timeav{\intBCnab 2q1  \vvvdspar}{\vvvlspar} 
 +\timeav{\intBCnab 1q1 \vvvdspar}{\vlsperp} 
 \\ \nn
 &+
 \edbLM{20.0}{\pardiv \timeav{\intBCnab 2q{\zs} \vvvdspar}{\ppperp \vvvlspar}} 
 +\timeav{\intBCnab 1q{\zs} \vvvdspar}{\ppperp\vlsperp}
 \Big\},
 \\
 \eqlab{vddFinalParOAppendix}
 \vvvddspar &= \frac{2}{\omega \delta^2} \Big\{
 -\pargrad \Big[\timeav{\intBCnab 2qq \vdsperp}{\vdsperp}
 \\ \nn
 &
 \qquad\qquad\qquad
 +\timeav{\intBCnab 21q \vlsperp}{\vdsperp}
 +\timeav{\intBCnab 2q1 \vdsperp}{\vlsperp}\Big]
 \\ \nn
 &
 +\pardiv \timeav{\intBCnab 2qq \vvvdspar}{\vvvdspar} 
 +\timeav{\intBCnab 1qq  \vvvdspar}{\vdsperp}  
 \\ \nn
 &+
 \pardiv \timeav{\intBCnab 21q \vvvlspar}{\vvvdspar}  
 +\timeav{\intBCnab 11q  \vvvlspar}{\vdsperp}  
 \\ \nn
 &+
 \pardiv \timeav{\intBCnab 2q1  \vvvdspar}{\vvvlspar} 
 +\timeav{\intBCnab 1q1 \vvvdspar}{\vlsperp} 
 \\ \nn
 &
 +\timeav{\intBCnab 1{\zs}q \ppperp  \vvvlspar}{\vdsperp} 
 +\timeav{\intBCnab 1q{\zs} \vvvdspar}{\ppperp\vlsperp}
 \Big\},
 \eal
 \bal
 \eqlab{vddOFinalPerpAppendix}
 \vddperpO &= \vddperppO + \vddperpvO = -\frac{2}{\omega \delta^2} \Big\{
 \\ \nn
 &
 +\pardiv\big[\pardiv \timeav{\intnab 3qq \vvvdspar}{\vvvdspar} 
 +\timeav{\intnab 2qq \vvvdspar}{\vdsperp}\big] 
 \\ \nn
 &
 +\pardiv \big[\pardiv \timeav{\intnab 31q \vvvlspar}{\vvvdspar}   %
 +\timeav{\intnab 21q \vvvlspar}{\vdsperp}\big] 
 \\ \nn
 &		
 +\pardiv\big[
 \edbLM{19.5}{\pardiv \timeav{\intnab 3{\zs}q \ppperp\vvvlspar}{\vvvdspar}} 
 +\timeav{\intnab 2{\zs}q \ppperp\vvvlspar}{\vdsperp} \big]  
 \\ \nn
 &
 +\pardiv \big[\pardiv \timeav{\intnab 3q1 \vvvdspar}{\vvvlspar} 
 + \timeav{\intnab 2q1 \vvvdspar}{\vlsperp} \big]  
 \\ \nn\
 &
 +\pardiv \big[
 \edbLM{19.5}{\pardiv \timeav{\intnab 3q{\zs} \vvvdspar}{\ppperp \vvvlspar}} 
 +\timeav{\intnab 2q{\zs} \vvvdspar}{\ppperp \vlsperp} \big]  
 \Big\}.
 \eal
 \esuba

%
%

\section{Implementation in COSMOL of the 2nd-order boundary condition}
\seclab{COMSOLimplem}
The basic COMSOL Multiphyiscs syntax for implementing the BL25 model is presented in  \secref{COMSOL_impl} together with the COMSOL implementation (written in the \qmtt{tt-typeface}) of the BL25 fluid-solid first-order  boundary conditions. To implement the more involved 2nd-order BL25 slip-velocity boundary condition \eqref{v20Final}, we need to provide COMSOL implementations for terms of the form (A) $\pargrad \timeav{\intBCnab 2qq \vdsperp}{\vdsperp}$, (B) $\pardiv \timeav{\intBCnab 2qq \vvvdspar}{\vvvdspar}$, and (C) $\timeav{\intBCnab 1q\zs  \vvvdspar}{\ppperp\vlsperp}$  from  \eqref{vddFinalPar}; (D) $\pardiv\pardiv \timeav{\intBCnab 3qq \vvvdspar}{\vvvdspar}$ and (E) $\pardiv \timeav{\intBCnab 2q1 \vvvdspar}{\vlsperp}$
from \eqref{vddFinalPerp}; and (F) $\timeav{\ii\vvvwall_1\cdot\:}{\grad \vvvls_1+\pargrad \vvvds_1} + \timeav{\vwallperp}{\ks\vvvds_1}$ from \eqref{StokesDriftWall}. In the following, we implement (A)-(F) term by term. However, we note that in expressions containing a parallel vector $\vvvdspar$, it is easier to implement the full vector $\vvvds_1$, and then project the final result onto the tangential plane spanned by the tangent vector $\ttt_1$ and $\ttt_2$. The procedure is exemplified as follows:
 \bsubal{tangentproj}
 \pardiv \timeav{\intBCnab 2qq \vvvdspar}{\vvvdspar} &= (\WWW\cdot\ttt_1)\:\ttt_1 + (\WWW\cdot\ttt_2)\:\ttt_2,
 \\
 \text{for }\; \WWW &= \pardiv \timeav{\intBCnab 2qq \vvvds_1}{\vvvds_1}.
 \esubal
In each  term, the tangential derivative $\grad_\parallel$ projects one of the velocity fields onto the tangential plane, but not the other. Hence the necessity of the final tangent-plane projection. In COMSOL, the $K$th component of the tangent-plane projection of $\WWW$ is given by the expression \qmtt{(WX*t1X+WY*t1Y+WZ*t1Z)*t1K + (WX*t2X+WY*t2Y+WZ*t2Z)*t2K}.

(A) Introducing the tangential divergence \qmtt{divparvd10 = vd1XtX+vd1YtY+vd1ZtZ}, $\vdsperp$ is given by \eqref{vdsperp} as \qmtt{vd1zeta = (i/ks)*divparvd10}. Next, with \qmtt{twoOVERomgdsqr} being the pre-factor $\frac{2}{\omega \delta^2}$, all factors $\intBCnab nab$ of \eqref{IabnBC} multiplied by $\frac{2}{\omega \delta^2}$ are called \qmtt{Jabn}, \eg, \qmtt{Jqz2 = twoOVERomgdsqr*(-(1-i)/2)*dvisc\^{}3}. With\hfill\qmttL{Jqq2vd1zetavd1zeta}\hfill\qmttC{=}\hfill\qmttC{1/2*realdot(Jqq2*}
\\
\qmttR{vd1zeta,vd1zeta)}, the $k$th
component of the tangential gradient $\frac{2}{\omega \delta^2}\pargrad \timeav{\intBCnab 2qq \vdsperp}{\vdsperp}$ is implemented as
\qmtt{dtang(Jqq2vd1zetavd1zeta,k)} followed by the tangent-plane projection~\eqnoref{tangentproj}.

(B) Since $\pardiv \timeav{\intBCnab 2qq \vvvdspar}{\vvvdspar} = \timeav{\intBCnab 2qq \vvvdspar}{\pardiv \vvvdspar} +  \timeav{\intBCnab 2qq \bm \nabla_\parallel \vvvdspar}{\cdot\vvvdspar}$, we introduce for K = X, Y, and Z the intermediate variable \qmtt{Jqq2vd1KDIVvd10 =} \qmttR{0.5*realdot(Jqq2*vd1K,divparvd10)} \hfill together \hfill with
\\
\qmttL{Jqq2DOTGRADvd1Kvd10 =} \hfill \qmttC{0.5*realdot(Jqq2*vd1KtX,}
\\
\qmttR{vd1X) + 0.5*realdot(Jqq2*vd1KtY,vd1Y)} \hfill \qmttC{+} \hfill \qmttC{0.5*}
\\
\qmttR{realdot(Jqq2*vd1KtZ,vd1Z))}. Consequently, the $K$th component of $\frac{2}{\omega \delta^2}\pardiv \timeav{\intBCnab 2qq \vvvdspar}{\vvvdspar}$  is implemented as
\qmtt{Jqq2vd1KDIVvd10 + Jqq2DOTGRADvd1Kvd10}  followed by the tangent-plane projection~\eqnoref{tangentproj}.

(C) Using \eqref{cont_1} and the tangential divergence, we find $\ppperp\vlsperp = \div\vvvls_1 - \pardiv\vvvls_1 =  \ii\omega\kapO p^0_1 - \pardiv\vvvls_1$, and therefore we introduce the variables \qmtt{divparvf10 = vf1XtX+vf1YtY+vf1ZtZ} and \qmtt{vf1zetazeta = i*kap0*omega*p1-divparvf10}. The $K$th component of $\frac{2}{\omega \delta^2}\timeav{\intBCnab 1q\zs  \vvvdspar}{\ppperp\vlsperp}$ is then simply implemented as \qmttR{0.5*realdot(Jqz1*vd1K,vf1zetazeta)}, followed by the tangent-plane projection~\eqnoref{tangentproj}.

(D) $\pardiv \pardiv \timeav{\intBCnab 3qq \vvvdspar}{\vvvdspar} = \pardiv \pardiv \timeav{\intBCnab 3qq \vvvds_1}{\vvvds_1}$, since the two tangential divergences ensure that the full 3D vectors are projected onto the tangent plane. Now, the $K$th component of the vector resulting after the first $\pardiv$ is
\qmttL{DIVJqq3vd1Kvd1 = 0.5*(dtang(realdot(Jqq3*vd1K,vd1X),x) +
dtang(}
\\
\qmttC{realdot(Jqq3*vd1K,vd1Y),y)+
dtang(realdot(Jqq3}
\\
\qmttR{*vd1K,vd1Z),z))}. So $\frac{2}{\omega \delta^2}\pardiv \pardiv \timeav{\intBCnab 3qq \vvvdspar}{\vvvdspar}$  is simply implemented as
\qmttL{dtang(DIVJqq3vd1Xvd1,x)+ dtang(DIVJqq3vd1Yvd1,y)+}\qmttR{dtang(DIVJqq3vd1Zvd1,z)}.

(E) $\pardiv \timeav{\intBCnab 2q1 \vvvdspar}{\vlsperp} = \pardiv \timeav{\intBCnab 2q1 \vvvds_1}{\vlsperp}$, since the tangential divergence ensures that $\vvvds_1$ is projected onto the tangent plane. The $K$th component of the argument to the divergence is implemented as
\qmttL{Jq12vd1Kvf1zeta} \qmttR{= 0.5*realdot(Jq12*vd1X,vf1zeta)}, and therefore,
$\frac{2}{\omega \delta^2}\pardiv\timeav{\intBCnab 2q1 \vvvdspar}{\vlsperp}$ is implemented simply as
\qmttL{dtang(}
\\
\qmttR{Jq12vd1Xvf1zeta,x) + dtang(Jq12vd1Yvf1zeta,y)+}
\qmttR{dtang(Jq12vd1Zvf1zeta,z)}.

(F) The Stokes drift contribution is the term $\frac{1}{\omega}\big[\timeav{\ii\vvvwall_1\cdot\:}{\grad \vvvls_1+\pargrad \vvvds_1} + \timeav{\vwallperp}{\ks\vvvds_1}\big]$. Its $K$th component\hfill is\hfill implemented\hfill as\hfill \qmttL{vSD2K = 0.5/omega*(}
\\
\qmttC{realdot(i*vs1X,vf1KX+vd1XtX)}\hfill \qmttC{+}\hfill \qmttC{realdot(i*vs1Y,}
\\
\qmttC{vf1KY+vd1XtY}\hfill \qmttC{+}\hfill \qmttC{realdot(i*vs1Z,vf1KZ+vd1XtZ) +}
\\
\qmttR{realdot(vs1zeta,ks*vd1K))}. The perpendicular com-\\ponent is given by
\qmttL{vSD2zeta = nX*vSD2X + nY*vSD2Y}
\\
\qmttR{+ nZ*vSD2Z0} and the $K$th component of the tangential projection
is \qmtt{vSD2parK = vSD2K-nK*vSD2zeta}.

The final BL25-model slip-velocity boundary condition~\eqnoref{v20Final} is obtained by adding all terms of the six basic forms (A)-(F). One form may have several contributions with different pre-factors \qmtt{Jabn}, $a,b = 1, q, \zeta$ and $n = 1,2,3$, and different velocity components.

\begin{figure*}[t]
\centering
\includegraphics[width=0.9\textwidth]{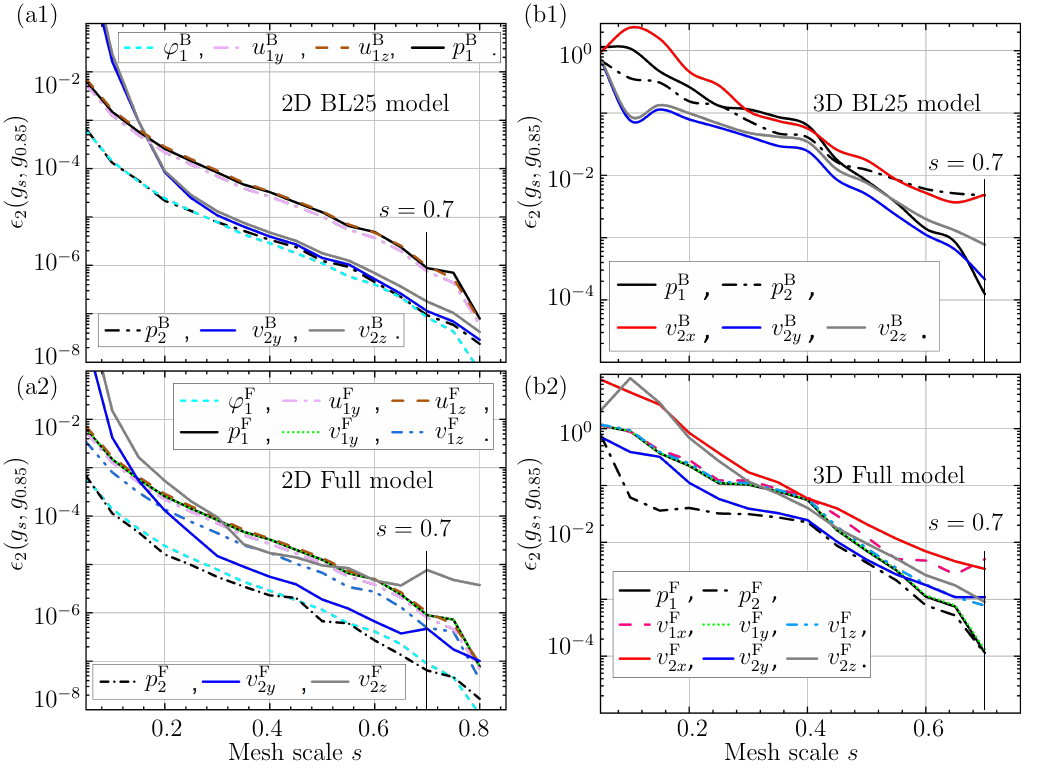}
\caption[]{\figlab{Mesh_conv}
Semilog mesh-convergence plot of the relative deviation $\eps^\notop_2(g_s,g_{0.85})$ defined in \eqref{L2norm} versus the scaling parameter~$s$.
(a1) BL25 model in 2D: field components in 1st order $\varphi_1^{\text{B}}$, $u_{1y}^{\text{B}}$, $u_{1z}^{\text{B}}$, and $p_1^{\text{B}}$, and in 2nd order $p_{2}^{\text{B}}$, $v_{2y}^{\text{B}}$, and $v_{2z}^{\text{B}}$.
(a2) Full model in 2D: field components in 1st order $\varphi_1^{\text{F}}$, $u_{1y}^{\text{F}}$, $u_{1z}^{\text{F}}$, $p_1^{\text{F}}$, $v_{1y}^{\text{F}}$, and $v_{1z}^{\text{F}}$,  and in 2nd order $p_2^{\text{F}}$, $v_{2y}^{\text{F}}$, and $v_{2z}^{\text{F}}$.
(b1) BL25 model in 3D: field components in 1st order  $p_1^{\text{B}}$, and in 2nd order $p_2^{\text{B}}$,
$v_{2x}^{\text{B}}$, $v_{2y}^{\text{B}}$, and $v_{2z}^{\text{B}}$.
(b2) Full model in 3D: field components in 1st order $p_1^{\text{F}}$, $v_{1x}^{\text{F}}$, $v_{1y}^{\text{F}}$, and in 2nd order $v_{1z}^{\text{F}}$, $p_2^{\text{F}}$, $v_{2x}^{\text{F}}$, $v_{2y}^{\text{F}}$ and $v_{2z}^{\text{F}}$.
}
\end{figure*}

%
%

\section{Mesh convergence study}
\seclab{MeshConvergence}

We have performed a standard mesh-convergence study \cite{Muller2012}. In 2D simulations, the solid PMMA, glue layer, and PZT domains are discretized using a free triangular mesh with maximum and minimum element size $h_{\text{max}}^\sl = 0.4H_f/s$ and $h_{\text{min}}^\sl=0.025h_{\text{max}}^\sl$, where $s$ is the scaling parameter that is gradually increased for refining the mesh. Since the electrode thickness is small, we have used the COMSOL mapped mesh with 40 as the number of elements for smoothing the mesh from the glue layer to the PZT. For the fluid domain, we defined maximum and minimum element size as $h_{\text{max}}^\fl = h_{\text{min}}^\sl$ and $h_{\text{min}}^\fl = 0.12h_{\text{max}}^\fl$. Further, we have used the boundary layer mesh to accurately capture velocity fields in the Full model. The boundary-layer field is defined across the fluid-solid interface using 8 layers, with the first layer of $0.2\delta$ and a stretching factor of $1.2$ so that we have in total depth of $3.4\delta$.

In the case of the  Full model in 3D, the solid PMMA domain is discretized using the free triangular mesh with maximum and minimum element size $h_{\text{max}}^\sl = 0.7H_f/s$ and $h_{\text{min}}^\sl = 0.125h_{\text{max}}^\sl$. Since the thin-film transducer is of thickness of $2~\SImum$, we discretized the transducer using the structured swept mesh with the swept directed from the top surface of the thin-film transducer. We discretized the fluid domain with a triangular mesh of maximum and minimum element size $h_{\text{max}}^\fl = h_{\text{min}}^\sl$ and $h_{\text{min}}^\fl = 0.6h_{\text{max}}^\fl$. Since a much finer mesh is required at the fluid-solid interface for the Full model to converge, we have used an edge mesh at the interface with maximum element size $h_{\text{max}}^\text{edge} = 0.45h_{\text{max}}^\fl$. Further, we have implemented the boundary-layer mesh using 8 layers to resolve the boundary-layer fields with the first-layer thickness set to $0.2\delta$.

For the BL25 model in 3D, we use a much coarser mesh inside the fluid domain and we do not need a boundary-layer mesh. Accordingly, the solid PMMA domain is discretized using the triangular mesh with maximum and minimum element size $h_{\text{max}}^\sl = 0.4H_f/s$ and $h_{\text{min}}^\sl = 0.2h_{\text{max}}^\sl$. Similar to the 3D Full model, we use a swept mesh for the thin-film transducer, and  coarser triangular mesh in the fluid domain with maximum and minimum element size $h_{\text{max}}^\fl = h_{\text{min}}^\sl$  and $h_{\text{min}}^\fl = 0.6h_{\text{max}}^\fl$. In \figref{Mesh_conv}, the relative deviation $\eps_2(\gnn_s,\gnn_{0.85})$, defined in \eqref{L2norm},  of the solution $\gnn_s$ with mesh scale $s$ from the solution $\gnn_{0.85}$ with the maximum mesh scale $s=0.85$ (the finest mesh),  is plotted in a semilog plot versus the mesh scale $s$ for both the 2D and the 3D simulation model.

%
%


%

\end{document}